\documentclass[aps,prb,twocolumn,superscriptaddress,floatfix]{revtex4-2}
\usepackage{qcircuit}
\usepackage{amsmath,bm}
\usepackage[dvips]{graphicx}
\usepackage{amsmath,amssymb,amsthm,mathrsfs,amsfonts,dsfont}
\usepackage{subfigure, epsfig}
\usepackage{braket}
\usepackage{ulem}
\usepackage{bm}
\usepackage{enumerate}
\usepackage{color}
\usepackage{graphicx}
\usepackage{algorithm}
\usepackage{algpseudocode}
\usepackage{comment}
\usepackage{hyperref}

\newcommand{\sbraket}[1]{\bigl[\hspace{0.5mm}{#1}\hspace{0.5mm}\bigr]}

\newcommand{\tr}{\mathrm{Tr}}

\begin{document}


\title{Error-Mitigated Quantum Metrology via Virtual Purification}


\author{Kaoru Yamamoto}
\email{kaoru.yamamoto.uw@hco.ntt.co.jp}
\affiliation{NTT Computer and Data Science Laboratories, NTT Corporation, Musashino 180-8585, Japan}

\author{Suguru Endo}
\email{suguru.endou.uc@hco.ntt.co.jp}
\affiliation{NTT Computer and Data Science Laboratories, NTT Corporation, Musashino 180-8585, Japan}
\affiliation{JST, PRESTO, 4-1-8 Honcho, Kawaguchi, Saitama, 332-0012, Japan}

\author{Hideaki Hakoshima}
\email{hakoshima.hideaki.qiqb@osaka-u.ac.jp}
\affiliation{Research Center for Emerging Computing Technologies, National Institute of Advanced Industrial Science and Technology (AIST), Central2, 1-1-1 Umezono, Tsukuba, Ibaraki 305-8568, Japan}
\affiliation{Center for Quantum Information and Quantum Biology, Osaka University, 1-2 Machikaneyama, Toyonaka 560-0043, Japan}

\author{Yuichiro Matsuzaki}
\email{matsuzaki.yuichiro@aist.go.jp}
\affiliation{Research Center for Emerging Computing Technologies, National Institute of Advanced Industrial Science and Technology (AIST), Central2, 1-1-1 Umezono, Tsukuba, Ibaraki 305-8568, Japan}
\affiliation{
NEC-AIST Quantum Technology Cooperative Research Laboratory,
National Institute of Advanced Industrial Science and Technology (AIST), Tsukuba, Ibaraki 305-8568, Japan
}
\author{Yuuki Tokunaga}
\email{yuuki.tokunaga.bf@hco.ntt.co.jp}
\affiliation{NTT Computer and Data Science Laboratories, NTT Corporation, Musashino 180-8585, Japan}


\begin{abstract}
Quantum metrology with entangled resources aims to achieve sensitivity beyond the standard quantum limit by harnessing quantum effects even in the presence of environmental noise. So far, sensitivity has been mainly discussed from the viewpoint of reducing statistical errors under the assumption of perfect knowledge of a noise model. However, we cannot always obtain complete information about a noise model due to coherence time fluctuations, which are frequently observed in experiments. Such unknown fluctuating noise leads to systematic errors and nullifies the quantum advantages. Here, we propose an error-mitigated quantum metrology that can filter out unknown fluctuating noise with the aid of purification-based quantum error mitigation. We demonstrate that our protocol mitigates systematic errors and recovers superclassical scaling in a practical situation with time-inhomogeneous bias-inducing noise.
Our result is the first demonstration to reveal the usefulness of purification-based error mitigation for unknown fluctuating noise, thus paving the way not only for practical quantum metrology but also for quantum computation affected by such noise. 
\end{abstract}

\maketitle

\textit{Introduction.---}
Quantum metrology with entangled resources has been shown to reach the Heisenberg limit of sensitivity with respect to the number of qubits \cite{giovannetti2004quantum, giovannetti2006quantum, giovannetti2011advances, toth2014quantum, degan2017quantum, pezze2018quantum,trenyi2022multicopy}. It may provide significant improvements for versatile applications such as atomic-frequency \cite{huelga1997improvement, leibfried2004toward} and electron-spin-resonance measurements \cite{budoyo2020electron,toida2019electron}, magnetometry \cite{balasubramanian2008nanoscale,maze2008nanoscale}, thermometry \cite{neumann2013high-precision}, and electrometers \cite{dolde2011electric}.
\begin{figure*}[ht]
    \centering
    \includegraphics[width= \linewidth]{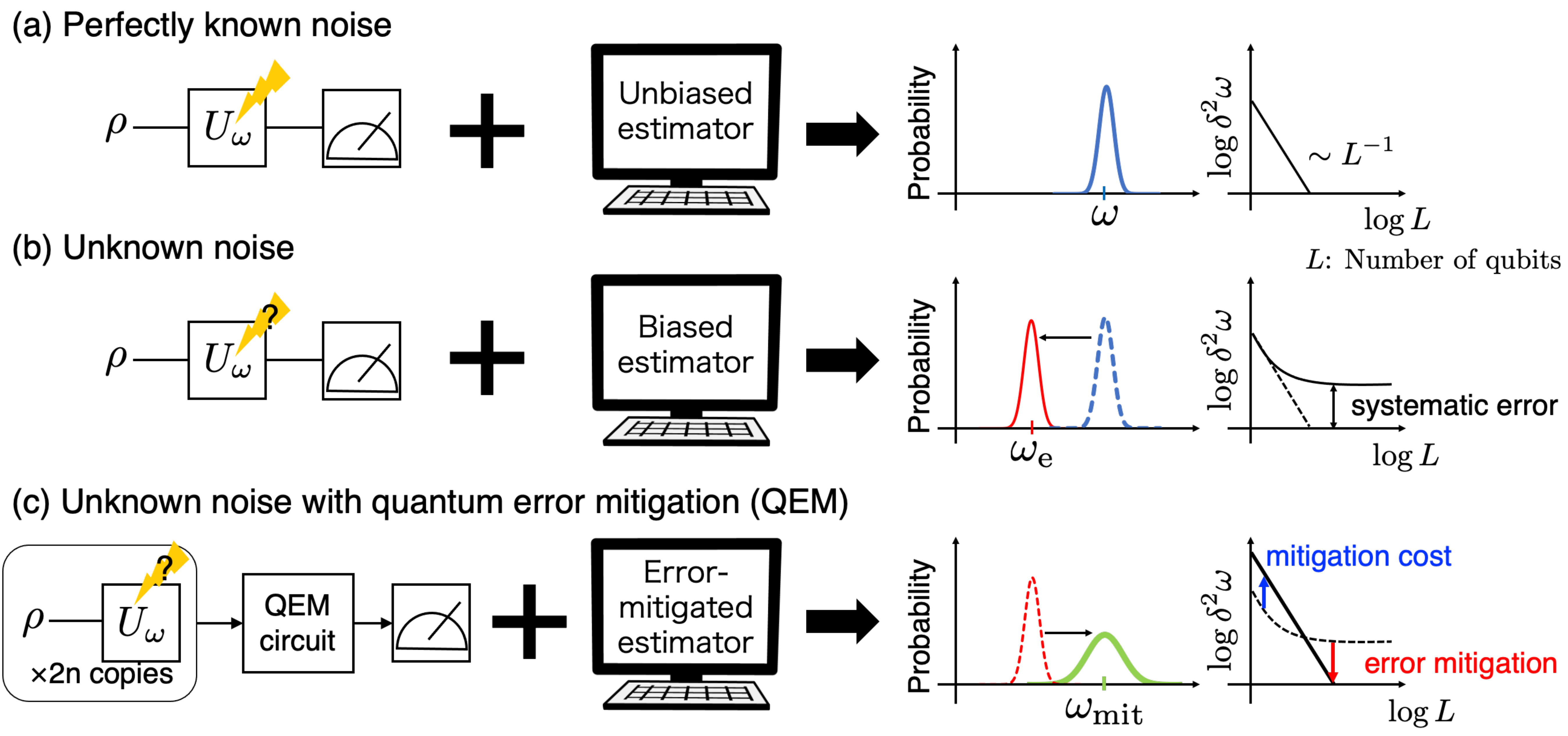}
    \caption{Schematic illustration of the present work. In the standard quantum metrology, we prepare the initial state $\rho$ composed of $L$ qubits, expose it to the target field described by the time-evolution operator $U_\omega$, where some noise process occurs, and measure the parametrized final state. 
    After many iterations, we obtain the average data. Separately, we theoretically calculate the estimator to estimate $\omega$ from the average data. (a) Even under noise, perfect knowledge of the noise model provides the unbiased estimator and leads to no systematic errors. (b) Imperfect knowledge of the noise model induced by e.g.~unknown coherence-time fluctuation leads to a biased estimator and involves systematic errors, thus resulting in deterioration of scaling of $\delta^2\omega$, the estimation uncertainty of $\omega$. (c) In the error-mitigated quantum metrology, $2n$ copies of the states after time evolution are put into QEM circuits. Then we calculate the error-mitigated estimator to estimate $\omega$ with the average data. Our protocol reduces systematic errors and recovers the scaling of $\delta^2\omega$. Note that the schematic picture of the scaling shows a case of Markovian noise.} 
   \label{fig:schematic}
\end{figure*}

The Heisenberg limit is susceptible to decoherence; for example, under the effect of Markovian dephasing, the sensitivity of entangled states scales to the standard quantum limit (SQL), as do separable states \cite{huelga1997improvement}.
Several theoretical studies predict that scaling beyond the standard quantum limit is possible: they include the superclassical scaling, which is also called Zeno scaling, under the time-inhomogeneous noise model \cite{matsuzaki2011magnetic, chin2012quantum}, quantum scaling by quantum teleportation \cite{averin2016suppression,matsuzaki2018quantum}, using the collective effect of open quantum systems \cite{kukita2021heisenberg,beau2017nonlinear}, and applying quantum error correction \cite{kessler2014quantum, dur2014improved}. So far, however, sensitivity scaling has been mainly discussed from the viewpoint of statistical errors under the assumption that the noise model can be fully characterized [Fig.~\ref{fig:schematic}(a)].

In experiments, we cannot always obtain complete information of a noise model typically due to coherence-time fluctuations \cite{muller2015interacting, yan2016the, abdurakhimov2016a, kandara2019Nature}; accordingly, noise characterization becomes intractable, leading to ``systematic errors'' [Fig.~\ref{fig:schematic}(b)].
Systematic errors usually result from a difference between the actual situation and the theoretical estimator used by experimentalists to estimate the target parameter. Intractable noise characterization leads to a biased estimator and induces systematic errors in estimations. In practice, systematic errors are fatal to quantum metrology because they cannot be reduced even when the number of qubits increases, thus seriously limiting any sensitivity improvement \cite{wolf2015subpicotesla,budoyo2020electron,rojkov2021bias}. Despite systematic errors typically being present in experiments, there is as yet no general approach to dealing with them, although some studies have tackled specific scenarios \cite{strubi2013measuring,pang2016protecting,martinez2016can,shimada2021quantum}.

In the present Letter, we propose a quantum-metrology protocol incorporating quantum error mitigation (QEM) to mitigate systematic errors, thereby improving the scaling of sensitivity even in the presence of unknown fluctuating noise [Fig.~\ref{fig:schematic}(c)]. While conventional QEM methods have been designed for suppressing systematic errors in the expectation values produced by near-term quantum algorithms \cite{endo2021hybrid,temme2017error,endo2018practical,li2017efficient,kandara2019Nature,song2019quantum,zhang2020error,mcardle2019error,bonet2018low,sun2021mitigating,larose2020mitiq}, they are not suitable for suppressing the systematic errors coming from unknown fluctuating noise. For example, probabilistic error cancellation cancels the effect of noise by inverting the noise map based on the characterization of the noise \cite{temme2017error,endo2018practical}, while error extrapolation assumes the precise control of the noise model \cite{temme2017error,li2017efficient}; thus, unknown fluctuating noise seriously degrades the performance of QEM. To deal with this problem, we construct the protocol of error-mitigated quantum metrology based on purification-based QEM \cite{koczor2020exponential,huggins2020virtual}, so that it can filter out unknown fluctuating noise that differs from one experimental run to another. It is noteworthy that our work is the first proposal to use purification-based QEM for overcoming unknown time-fluctuating noise. In numerical simulations, we used it to suppress bias-inducing Markovian and time-inhomogeneous noise and thereby restore the scaling of sensitivity with respect to the number of qubits. In particular, we observed the superclassical scaling with our method.

\textit{Quantum metrology with systematic errors.---}
Here, we describe a general theory for systematic errors in a Ramsey-type measurement of quantum metrology \cite{sugiyama2015precision,takeuchi2019quantum,okane2020quantum}.
In a typical quantum-metrology setup, we prepare an initial state, expose this state to the target fields characterized by a parameter $\omega$, and obtain a state $\rho$.
Then, we perform a measurement on this state that can be described by a projection operator $P$ producing a binary outcome. The measurement outcome, $m_j \in \{-1,1\}$, is obtained from the measurement probability,
\begin{equation}
p = \tr[P\rho] = x+y\omega, \label{eq:P_c}
\end{equation}
where $x$ and $y$ are some scalars.
Here, we have assumed that $\omega$ is small and have ignored the higher order terms of $\omega$. Repeating the measurement $N_\text{samp}$ times yields the average value of data, $S_{N} = \sum_{j=1}^{N_\text{samp}}m_j/N_\text{samp}$. To estimate the parameter $\omega$, we need to fit this experimental data with a theoretical estimator. 
To obtain the estimator, we first consider the theoretical density matrix $\rho_\text{e}$ and calculate the estimated probability as
\begin{equation}
p_\text{e} = \tr[P\rho_\text{e}] = x_\text{e}+y_\text{e}\omega, \label{eq:P_e}
\end{equation}
where, $x_\text{e}$ and $y_\text{e}$ denote the \textit{estimated} values of $x$ and $y$, respectively. By reference to Eq.~\eqref{eq:P_e}, we calculate the estimator of $\omega$ as $\omega_\text{e} = (S_{N}-x_\text{e})/y_\text{e}$ and estimate $\omega$ using the average data $S_N$ with the estimator.
If we have imperfect knowledge of noise model, $\rho_\text{e}$ would be different from the true one and leads to a biased estimator, thereby leading to systematic errors.

Systematic errors require us to consider the estimation uncertainty of the target quantity \cite{sugiyama2015precision,takeuchi2019quantum,okane2020quantum}. The estimation uncertainty of $\omega$ is defined as $\delta^2\omega = \Braket{(\omega-\omega_\text{e})^2}$, with the brackets denoting the ensemble average, and is calculated as
\begin{equation}
\delta^2 \omega = \frac{1}{y_\text{e}^2}\left[\text{Var}[p] + (x-x_\text{e})^2\right], \label{eq:delta^2omega}
\end{equation}
where $\text{Var}[p]$ is the variance of $p$, which is typically $\text{Var}[p] = p(1-p)/N_\text{samp}$, and we have neglected the higher order terms of $\omega$ because $\omega$ is small. Most of the previous theoretical studies focused on the first term in Eq.~\eqref{eq:delta^2omega} by assuming $\rho = \rho_\text{e}$, which comes from the statistical error, as it decreases with increasing $N_\text{samp}$. The second term in Eq.~\eqref{eq:delta^2omega} comes from the systematic error $x-x_\text{e}$ induced by the incorrect estimation of the probability $p\neq p_\text{e}$. Since $x-x_\text{e}$ remains even when $N_\text{samp}$ increases, it spoils the scaling of $\delta^2\omega$. In the following, we focus on reducing the systematic error $x-x_\text{e}$ by using QEM.

\begin{figure}
    \centering
    \includegraphics[width=\linewidth]{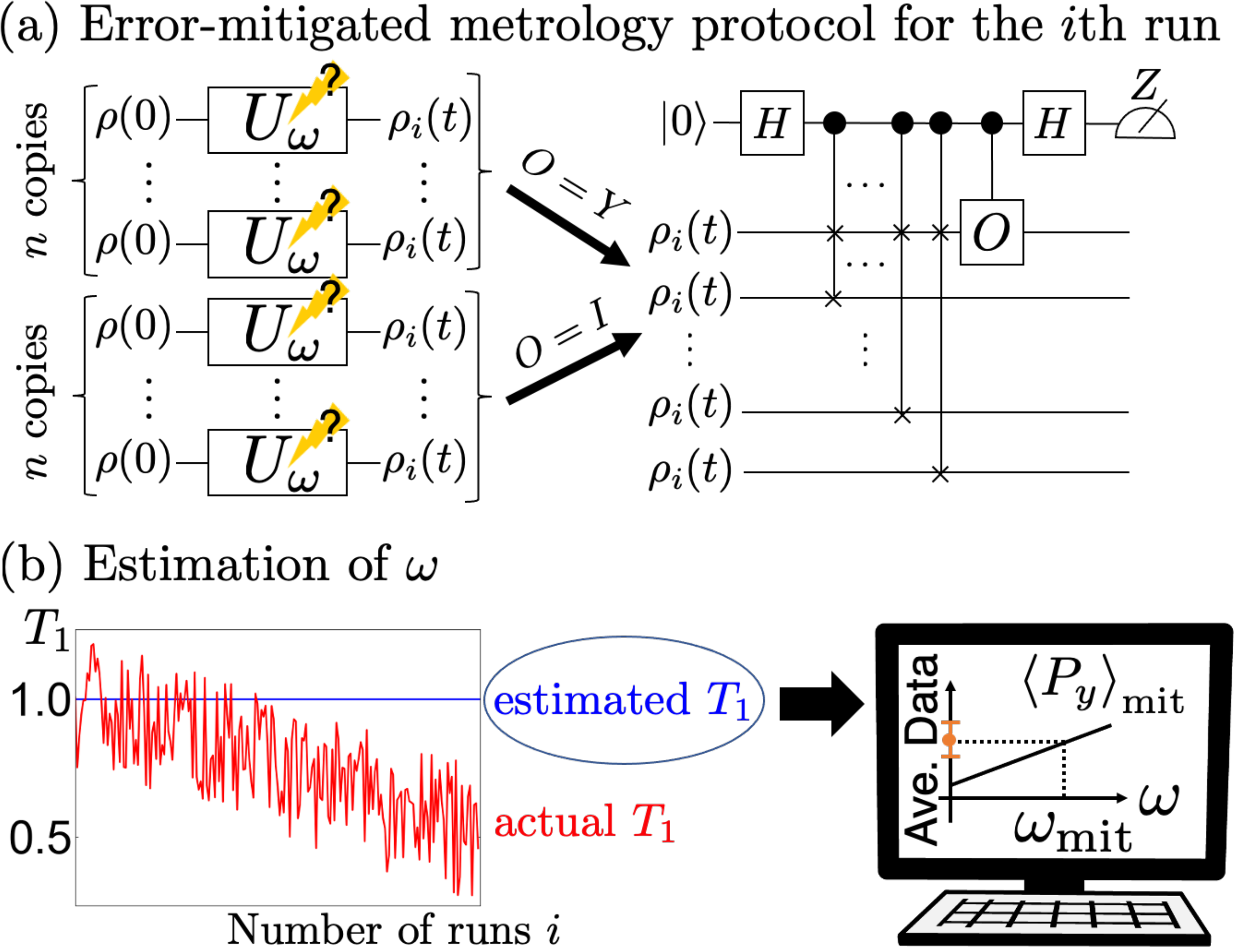}
    \caption{(a) Error-mitigated metrology protocol for the $i$th run. $2n$ copies of the initial states are exposed to the magnetic field, and $n$ copies are put to purification circuit \cite{koczor2020exponential} with $O=Y$ and the remains are put with $O=I$. (b) Schematic illustration of estimating $\omega$. Using estimated $T_1$, we calculate the error-mitigated estimator $\braket{P_y}_\text{mit} = \tr\left[\overline{\rho(t)^n} P_y\right]/\tr\left[\overline{\rho(t)^n}\right]$ and estimate $\omega$ with the data obtained in (a).}
   \label{fig:Numerics}
\end{figure}

\textit{Error-mitigated quantum metrology.---} 
Here, we introduce our general framework of error-mitigated quantum metrology, which is inspired by purification-based QEM \cite{koczor2020exponential,huggins2020virtual}. We assume that we implement $N_\text{samp}$ experimental runs and the noise fluctuates from one experimental run to another, i.e., we assume a different quantum state for the $i$th measurement described by $\rho_i$ in the $i$th experimental run. The key quantity of our framework is the following error-mitigated expectation value of the observable $O$ measured in the quantum circuit in Fig.~\ref{fig:Numerics}(a):
\begin{equation}
\braket{O}_\text{mit} = \frac{{\tr\bigl[\overline{\rho ^n}O\bigr]}}{\tr{\sbraket{\overline{\rho^n}}}},
\label{Eq:therem1}
\end{equation}
where $\overline{\rho^n}=\frac{1}{N_{\text{samp}}}\sum_{i=1}^{N_\text{samp}} (\rho_i)^n$ is the unnormalized purified density matrix made from $n$ copies of noisy density matrices; later, we will discuss its filtering effect on fluctuating noise.

Our protocol is described as follows [Figs.~\ref{fig:schematic}(c) and \ref{fig:Numerics}]: (1) make $2n$ copies of the initial state $\rho(0)$; (2) expose these $2n$ copies to the target field simultaneously (or almost at the same time) for an interaction time $t$, so that we can obtain the $2n$ copies of $\rho_i(t)$ even with fluctuating noise; (3) divide these $2n$ copies of $\rho_i(t)$ in half; (4) input $n$ of $2n$ density matrices to the purification circuit and obtain a ``single shot'' measurement outcome to calculate the numerator in Eq.~\eqref{Eq:therem1}; (5) input the remaining $n$ density matrices in a similar manner to (4) but with setting $O=I$, to calculate the denominator in Eq.~\eqref{Eq:therem1} \cite{procedure}; (6) repeat (1)--(5) and average the obtained data, $S_\text{num}$ for ${\tr\bigl[\overline{\rho ^n}O\bigr]}$ and $S_\text{denom}$ for $\tr{\sbraket{\overline{\rho^n}}}$: then, compute $S_{N_\text{mit}} \equiv S_\text{num}/S_\text{denom}$ as the estimator for $\braket{O}_\text{mit}$
; (7) calculate the error-mitigated estimator for $\omega$ as $\omega_\text{e} = (S_{N_\text{mit}}-x_\text{mit})/y_\text{mit}$ by calculating $\braket{O}_\text{mit} \equiv x_\text{mit}+y_\text{mit}\omega$ using estimated $\rho_\text{e}$, and then estimate $\omega$.  
Note that the density matrix for the ensemble of $N_\text{samp}$ states input to the purification circuit can be described as $\rho_\text{in}= \frac{1}{N_\text{samp}} \sum_{i=1}^{N_\text{samp}} \rho_i^{\otimes n}$; we can then obtain Eq.~\eqref{Eq:therem1} from a simple calculation \cite{supple}.

Now, we show that our protocol can filter out the noisy states and extract a dominant pure state even in the presence of fluctuating noise. Denoting the spectral decomposition of $\rho_i$ as $\rho_i=\sum_k p_k^{(i)} \ket{\psi_k^{(i)}}\bra{\psi_k^{(i)}}$, we have 
\begin{align}
\overline{\rho^n} &= \frac{1}{N_{\text{samp}}} (p_{\max})^n \sum_{ik} \left(\frac{p_k^{(i)}}{p_{\max}} \right)^n \ket{\psi_k^{(i)}}\bra{\psi_k^{(i)}} \notag \\
&= \frac{(p_{\max})^n}{N_{\text{samp}}} \ket{\psi_{\max}}\bra{\psi_{\max}}~(n \rightarrow \infty), 
\end{align}
where $p_{\max}= \max_{i,k}p_{k}^{(i)}$ and $\ket{\psi_{\max}}$ is the corresponding eigenstate. Thus, the contribution of states other than $\ket{\psi_{\max}}$ is exponentially suppressed as the number of copies $n$ increases, which means that our method can filter out unknown fluctuating noise. In general, the dominant eigenvector of the mixed state is distorted by noise and differs from the ideal quantum state, which is called coherent mismatch \cite{koczor2021dominant}. Nevertheless, our method clearly eliminates the systematic errors, to allow for a dramatic improvement in $\delta^2\omega$ in practical scenarios, as we will see later in the numerical simulations.

\textit{Demonstration of error-mitigated quantum metrology.---}
We demonstrate that our protocol mitigates systematic errors even under unknown fluctuating noise and improves the scaling of $\delta^2\omega$ using entanglement quantum metrology. Here, we consider Markovian and time-inhomogeneous local amplitude damping.
We choose the initial probe state $\rho(0)=\ket{\text{GHZ}}\bra{\text{GHZ}}$ as the $L$-qubit Greenberger-Horne-Zeilinger (GHZ) state $\ket{\text{GHZ}} =( \ket{0...0}+\ket{1...1})/\sqrt{2}$, where $\ket{0}$ and $\ket{1}$ are the eigenstates of the Pauli Z operator $\sigma_z$: $\sigma_z\ket{1} = \ket{1}$ and $\sigma_z\ket{0} = -\ket{0}$. We consider a uniform magnetic field described by the Zeeman Hamiltonian $H= \sum_{j=1}^{L}\omega \sigma_z^{(j)}/2$ with a parameter $\omega$ determined by the target field. Throughout the present Letter, we set $\hbar=1$ and assume small $L\omega t$. We also assume that the time needed for state preparation, error mitigation, and readout is much shorter than the interaction time with the magnetic fields. 
We consider that in the $i$th experimental run, the local amplitude damping with different error rates, $\epsilon_i(t)$, affects each state of the $2n$ copies in the time evolution. The state after the time evolution is described as $\rho_i(t)= \mathcal{E}(\epsilon_i)[e^{-iHt}\rho(0)e^{iHt}]$. Here, $\mathcal{E}(\epsilon_i)$ is the error map denoting the local amplitude damping, in which a single-qubit amplitude damping described by the Kraus operators, $K_{1}(\epsilon_i)= \ket{0}\bra{0} +  \sqrt{1-\epsilon_i(t)}\ket{1}\bra{1}$ and $K_{2}(\epsilon_i)= \sqrt{\epsilon_i(t)}\ket{0}\bra{1}$, acts on every $L$ qubit. The effect of fluctuating noise is included in $\epsilon_i(t) = 1-\exp(-t/T_{1,i})$ for Markovian noise and in $\epsilon_i(t) = 1-\exp[-(t/T_{1,i})^2]$ for time-inhomogeneous noise, where $T_{1,i}$ is the coherence time in the $i$th experimental run; we consider the actual coherence time drifting from $1.0$ to $0.5$ as $T_{1,i} = 1.0-0.5i/N_\text{samp}$ with uniform random fluctuation between $[-0.25, 0.25]$, while the estimated coherence time is assumed to be $T_{1,\text{e}}=1.0$ [Fig.~\ref{fig:Numerics}(b)]. We also fix the total experimental time as $T = N_\text{samp}t=100$.
From the $2n$ copies of $\rho_i(t)$, we obtain $\tr\left[\overline{\rho(t)^n}\right]$ with $O=I$ and $\tr\left[\overline{\rho(t)^n} Y\right]$ with $O=Y$, as shown in Eq.~\eqref{Eq:therem1}, where $Y = 2{P}_y-I$, $P_y = \ket{\text{GHZ}_y}\bra{\text{GHZ}_y}$, and $\ket{\text{GHZ}_y} = (\ket{0...0}-i\ket{1...1})/\sqrt{2}$. These expectation values lead to an error-mitigated probability of $\braket{P_y}_\text{mit}= \tr\left[\overline{\rho(t)^n} P_y\right]/\tr\left[\overline{\rho(t)^n}\right]$: see the Supplementary Materials (SM) for detailed calculations \cite{supple}. 

\begin{figure}
    \centering
    \includegraphics[width=0.9\linewidth]{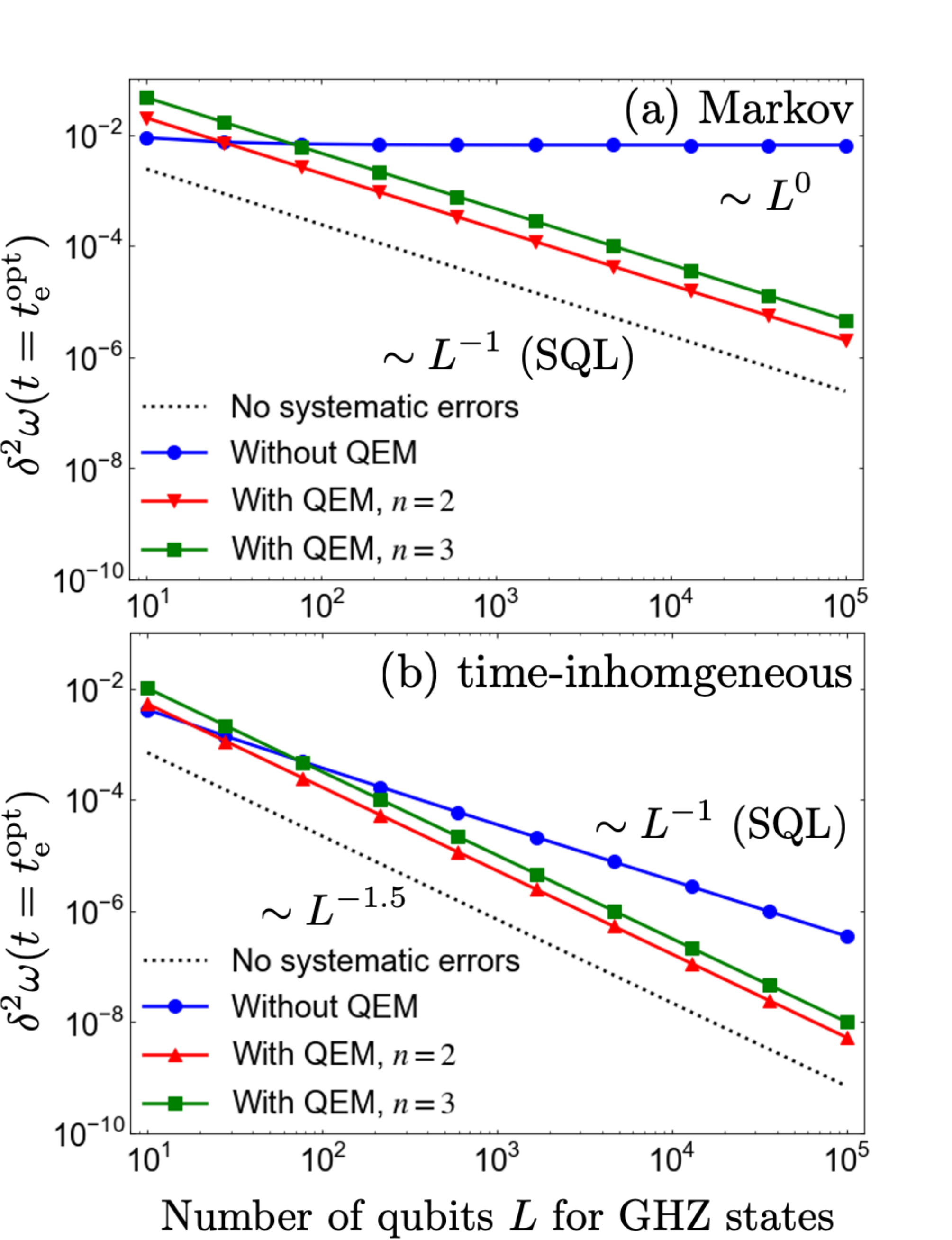}
    \caption{Estimation uncertainty $\delta^2\omega$ for (a) Markovian and (b) time-inhomogeneous local amplitude damping. Although the ideal scaling is realized by the unbiased estimator with the correct estimation of the noise model (black dotted line), a biased estimator with an incorrect estimation of the noise model leads to systematic errors and spoils the scaling (blue line with circles). Our protocol mitigates systematic errors and recovers the ideal scaling (red and green lines with triangles and squares, respectively).}
   \label{fig:VD}
\end{figure}

We now turn our attention to the numerical results.
The performance of our protocol is evaluated by comparing $\delta^2 \omega$ with and without QEM; see the numerical details in SM \cite{supple}.
The ideal case is shown by the black dotted line in Figs.~\ref{fig:VD}(a) and (b), when we consider that the actual coherence time is constant and correctly estimated, $T_{1,i} = T_{1,\text{e}}= 1.0$: $\delta^2\omega$ follows the conventional SQL scaling for Markovian noise, $\delta^2\omega \sim L^{-1}$, and superclassical scaling for time-inhomogeneous noise, $\delta^2\omega \sim L^{-1.5}$  \cite{matsuzaki2011magnetic,chin2012quantum}. However, when the actual coherence time is drifted as described above, the systematic error occurs and significantly spoils the scaling of $\delta^2\omega$ as shown by the blue line with points in Figs.~\ref{fig:VD}(a) and (b): $\delta^2\omega \sim L^{0}$ for Markovian noise and $\delta^2\omega \sim L^{-1}$ for time-inhomogeneous noise. The crucial reason for this deterioration is that the systematic error is not reduced by increasing $L$ for both cases as $|x-x_\text{e}| \sim L^0$, while the statistical error is reduced. Thus, for large enough $L$, the second term in Eq.~\eqref{eq:delta^2omega} coming from the systematic error is dominant and spoils the scaling. 

Now, we show that our protocol mitigates the systematic error and dramatically improves the scaling of $\delta^2\omega$. In Figs.~\ref{fig:VD}(a) and (b), the red line with triangles (for $n=2$) and the green line with squares (for $n=3$) show $\delta^2\omega$ in our error-mitigated quantum metrology; they demonstrate that our protocol recovers the SQL scaling for Markovian noise and superclassical scaling for time-inhomogeneous noise even when the estimated coherence time is different from the actual one. This demonstrates that our protocol successfully mitigates systematic errors and recovers the scaling; we also demonstrate the usefulness of our protocol for generalized amplitude damping in SM \cite{supple}.
In the purification-based QEM, as we increase the number of copies $n$, systematic errors are reduced better with an exponentially higher sampling cost \cite{koczor2020exponential}.
Therefore, the increase of $n$ improves (reduces)  the sensitivity when the systematic (statistical) error is dominant. Accordingly, we see the crossover between $n=1$ and $n\geq 2$ in Figs.~\ref{fig:VD}(a) and (b).
We also investigate how the number of copies of the GHZ states affects the uncertainty $\delta^2\omega$ to obtain the quantum enhancement in SM \cite{supple}.

Considering the trade-off between mitigating systematic error and increasing the statistical error, one may expect that there should be a crossover point between $n=2$ and $n=3$; indeed we can see it in the plot for another model in Fig.~\ref{fig:NV} \cite{supple}. In the present example in Fig.~\ref{fig:VD}, however, there is no crossover point between different $n$ for $n\geq 2$, and thus $n=2$ is always optimal for large enough $L$. This is because the systematic error decays as fast as or faster than the statistical error by increasing $L$ for $n\geq 2$; the former scales as $L^{-(n-1)}$ ($L^{-n}$), while the later scales as $L^{-1}$ ($L^{-3/2}$) for Markovian (time-inhomogeneous) noise. 
Therefore, the statistical error is always dominant for large enough $L$ for $n\geq 2$, and $n=2$ is optimal due to its least statistical error among $n\geq 2$.
This phenomenon comes from the $L$ dependence of the systematic error for $n\geq 2$ in the present example, $|x_{\text{e}}- x| \sim L^{-(n-1)}$ \cite{supple}.
We can relate the $L$ dependence of $|x-x_{\text{e}}|$ to the R\'enyi entropy of the error states \cite{supple}, which is related to the performance of purification-based QEM \cite{koczor2020exponential}.
In general, $L$ entangled qubits provide an exponentially increasing state space, and this typically increases the R\'enyi entropy of error states.
Thus, we expect that our protocol will work very well in quantum metrology using entangled states: see also the demonstrations of various models with different R\'enyi entropy in SM \cite{supple}.

\textit{Conclusion and outlook.---}
We proposed an error-mitigated quantum metrology to reduce systematic errors coming from an incorrect estimation of unknown noise typically induced by coherence-time fluctuations. Our error-mitigated quantum metrology, inspired by purification-based QEM, filters out fluctuating noise that differs from one experimental run to another. We used our method to suppress bias-inducing Markovian and time-inhomogeneous noise, where systematic errors spoil scaling of $\delta^2\omega$, and demonstrated restoration of the scaling. In particular, for the latter case, our method led to superclassical scaling. Here, we should mention that the number of copies of the input density matrix in our method may be reduced by using other methods related to purification-based QEM \cite{czarnik2021qubit,huo2021dual, cai2021resource} and that coherent error may be further reduced by combining our method with generalized subspace expansion \cite{yoshioka2021generalized}. 
Our results suggest that our scheme would be useful not only for quantum metrology affected by unknown fluctuating noise but also for quantum computation affected by such noise.

We thank Yang Wang for useful discussions. This work was supported by MEXT Q-LEAP Grant No. JPMXS0120319794; JST, PRESTO, Grants No. JPMJPR1919 and JPMJPR2114, Japan; JST COI-NEXT program (JPMJPF2014); JST [Moonshot R\&D] Grant No.~JPMJMS2061.
This Letter was (partly) based on results obtained from a project, JPNP16007, commissioned by the New Energy and Industrial Technology Development Organization (NEDO), Japan.

\bibliographystyle{apsrev4-2}
\bibliography{bib}

\begin{thebibliography}{66}
\bibitem [{pro()}]{procedure}%
  \BibitemOpen
  \href@noop {} {}\bibinfo {note} {Procedures (4) and (5) can be done in any
  order.}\BibitemShut {Stop}%
\end{thebibliography}


\clearpage
\widetext
\begin{center}
\textbf{\large Supplementary materials for ``Error-Mitigated Quantum Metrology via Virtual Purification''}
\end{center}
\setcounter{equation}{0}
\setcounter{figure}{0}
\setcounter{table}{0}
\setcounter{page}{1}
\makeatletter
\renewcommand{\theequation}{S\arabic{equation}}
\renewcommand{\thefigure}{S\arabic{figure}}
\renewcommand{\bibnumfmt}[1]{[S#1]}

In the Supplementary Materials, we provide a formulation for fluctuating noise in Sec.~\ref{sec:treat}, which is used in the present work. In Sec.~\ref{sec:Renyi}, we discuss the relationship between the systematic error and the R\'enyi entropy of the error states under a general incoherent noise. In Sec.~\ref{sec:NV}, we provide details of the calculations and discussions for the main text. The remains demonstrate the performance of our protocol for other models from the viewpoint of the R\'enyi entropy of the error states. In Sec.~\ref{sec:geneAD}, we consider a more generalized noise model than in the main text, a local amplitude damping at finite temperatures. In Sec.~\ref{sec:globalDepo}, we consider a global depolarizing noise on GHZ initial states, which involves almost the largest R\'enyi entropy. In Sec.~\ref{sec:NV}, we consider a model describing the nitrogen-vacancy center suffered by an energy relaxation in the high-temperature limit. Both models show the improvement of the sensitivity using our error-mitigated quantum metrology, thus suggesting the usefulness of our protocol.

\section{Treatment of fluctuating noise} \label{sec:treat}
Noise typically fluctuates, leading to a different density matrix for each measurement. Here, we explain how our protocol treats a different density matrix for each measurement induced by fluctuating noise. In the following, we denote the density matrix for the $i$th measurement as $\rho_i$. Let $X_i$ denote a random variable describing the $i$th measurement outcome in the quantum circuit in Fig.~\ref{fig:schematic}(c). The key assumption is that we can define the average and variance of $X_i$ using $\rho_i$ as follows:
\begin{align}
E[X_i] &= \frac{1+\text{Tr}\left[(\rho_i)^n O\right]}{2}, \\
\text{Var}[X_i] &= E[X_i]\left(1-E[X_i]\right) = \frac{1-\left\{\text{Tr}\left[(\rho_i)^n O\right]\right\}^2}{4}.
\end{align}
Introducing the average of $X_i$ as
\begin{equation}
\overline{X} = \frac{1}{N_\text{samp}}\sum_{i=1}^{N_\text{samp}} X_i
\end{equation}
with $N_\text{samp}$ being the number of samples, we can calculate the expectation value and variance of $\overline{X}$ as
\begin{align}
\text{E}\Bigl[\overline{X}\Bigr] =\text{E}\left[\frac{1}{N_\text{samp}}\sum_{i=1}^{N_\text{samp}} X_i\right] = \frac{1}{N_\text{samp}}\sum_{i=1}^{N_\text{samp}} \text{E}[X_i] = \frac{1}{N_\text{samp}}\sum_{i=1}^{N_\text{samp}} \frac{1+\tr\left[(\rho_i)^n O\right]}{2} = \frac{1+\tr\sbraket{\overline{\rho^n} O}}{2}, \label{eq:expect}
\end{align}
and 
\begin{align}
\text{Var}\sbraket{\overline{X}} =\text{Var}\left[\frac{1}{N_\text{samp}}\sum_{i=1}^{N_\text{samp}}X_i\right] 
= \frac{1}{(N_\text{samp})^2}\sum_{i=1}^{N_\text{samp}} \text{Var}[X_i] = \frac{1}{(N_\text{samp})^2}\sum_{i=1}^{N_\text{samp}} \frac{1-\left\{\tr\left[(\rho_i)^n O\right]\right\}^2}{4} = \frac{1-\overline{(\tr[\rho^n O])^2}}{4},\label{eq:variance}
\end{align}
where $\overline{A} = \sum_{i=1}^{N_\text{samp}}A_i/N_\text{samp}$ and we have assumed the independence of the $X_i$s.

\textit{The case of unitary $O$.---} 
The above $\text{E}\Bigl[\overline{X}\Bigr]$ is indeed the expectation value of the measurement in the error-mitigation circuit in Fig.~\ref{fig:schematic}(c); we obtain $p_\text{num} = \left(1+\tr\left[\overline{\rho^n} O\right]\right)/2$ and $\text{Var}[p_\text{num}] = \left(1-\overline{{(\tr[\rho^n O])^2}}\right)/4 $ as well as $p_\text{denom} = (1+\tr\sbraket{\overline{\rho^n}})/2$ and $\text{Var}[p_\text{num}] = \left(1-\overline{(\tr[\rho^n])^2}\right)/4$ with $O=I$. The error-mitigated expectation value of $O$ is calculated as
\begin{equation}
\braket{O}_\text{mit} = \frac{2p_\text{num}-1}{2p_\text{denom}-1} = \frac{{\tr\bigl[\overline{\rho ^n}O\bigr]}}{\tr\sbraket{\overline{\rho^n}}}.
\end{equation}

We can calculate the variance of $\braket{O}_\text{mit}$ by using error propagation as follows. Suppose we would like to calculate the variance of $z = f/g$. Then, we have
\begin{equation}
z = z_0+\Delta z = \frac{f_0+\Delta f}{g_0+\Delta g} \simeq \frac{f_0}{g_0}+ \frac{\Delta f}{g_0}-\frac{f_0\Delta g}{g_0^2},
\end{equation}
where $z_0$, $f_0$, and $g_0$ are the expectation values of $z$, $f$, and $g$, respectively, and $\Delta z \equiv z-z_0$, $\Delta f \equiv f-f_0$, and $\Delta g \equiv g-g_0$. The variance of $z$ is calculated as
\begin{equation}
\text{Var}[z] = \text{E}\left[(\Delta z)^2\right] = \frac{\text{Var}[f]}{g_0^2}+\frac{f_0^2\text{Var}[g]}{g_0^4}-2\frac{f_0\text{E}[\Delta f\Delta g]}{g_0^3}. \label{eq:varz}
\end{equation}
Inserting $f = 2p_\text{num}-1$ and $g=2p_\text{denom}-1$ into Eq.~\eqref{eq:varz} gives the variance of $\braket{O}_\text{mit}$ as
\begin{align}
\text{Var}\left[\braket{O}_\text{mit}\right]
&= \frac{1-\overline{\left(\text{Tr}[\rho^nO]\right)^2}}{N_\text{samp}\left(\text{Tr}\sbraket{\overline{\rho^n}}\right)^2} + \frac{\left(\overline{\text{Tr}[\rho^nO]}\right)^2\left\{1-\overline{\left(\text{Tr}[\rho^n]\right)^2} \right\}}{N_\text{samp}\left(\text{Tr}\sbraket{\overline{\rho^n}}\right)^4}, \label{eq:VarO}
\end{align}
where we have used $\text{Var}[f] = 4\text{Var}[p_\text{num}] = 1-\overline{(\tr\bigl[\rho^n O\bigr])^2}$ and $\text{Var}[g] = 4\text{Var}[p_\text{denom}] = 1-\overline{(\tr\bigl[\rho^n \bigr])^2}$. Here, we have assumed that there is no correlation between $p_\text{num}$ and $p_\text{denom}$, which leads to no correlation between $f = 2p_\text{num}-1$ and $g=2p_\text{denom}-1$ and $\text{E}[\Delta f\Delta g]=0$ in Eq.~\eqref{eq:varz}.

\textit{The case in the main text.---} 
However, the above calculation can not be directly applied to the case in the main text, where we would like to calculate $\braket{P_y}_\text{mit}$ but $P_y$ is not a unitary operation. In the main text, therefore, we set $O = Y$ defined by $P_y = (I+Y)/2$, which is unitary. Accordingly, the error-mitigation circuit in Fig.~\ref{fig:schematic}(c) provides $p_\text{num} = \left(1+\text{Tr}\bigl[\overline{\rho^n} Y\bigr]\right)/2$ and $p_\text{denom} = \left(1+\text{Tr}\sbraket{\overline{\rho^n}}\right)/2$. The error-mitigated expectation value of $P_y$ is then calculated to be
\begin{equation}
\braket{P_y}_\text{mit} = \frac{\text{Tr}\sbraket{\overline{\rho^n}} + \text{Tr}\bigl[\overline{\rho^n}Y\bigr]}{2\text{Tr}\sbraket{\overline{\rho^n}}}  = \frac{p_\text{denom}+p_\text{num}-1}{2p_\text{denom}-1}.
\end{equation}
Inserting $f = p_\text{denom}+p_\text{num}-1$ and $g=2p_\text{denom}-1$ into Eq.~\eqref{eq:varz} gives the variance of $\braket{P_y}_\text{mit}$ as
\begin{align}
\text{Var}\left[\braket{P_y}_\text{mit}\right]
&= \frac{2-\overline{(\text{Tr}[\rho^n])^2}-\overline{(\text{Tr}[\rho^nY])^2}}{4\left(\text{Tr}\sbraket{\overline{\rho^n}}\right)^2N_\text{samp}}  
+ \frac{\left(\text{Tr}\sbraket{\overline{\rho^n}}+ \text{Tr}\bigl[\overline{\rho^n}Y\bigr]\right)^2\left[1-\overline{(\text{Tr}[\rho^n])^2}\right]}{4\left(\text{Tr}\sbraket{\overline{\rho^n}}\right)^4N_\text{samp}} \notag \\
&-\frac{\left(\text{Tr}\sbraket{\overline{\rho^n}}+ \text{Tr}\bigl[\overline{\rho^n}Y\bigr]\right)\left[1-\overline{(\text{Tr}[\rho^n])^2}\right]}{2\left(\text{Tr}\sbraket{\overline{\rho^n}}\right)^3N_\text{samp}}, \label{eq:VarP_y}
\end{align}
where we have used $\text{Var}[f] = \text{Var}[p_\text{num}]+\text{Var}[p_\text{denom}] = \left[2-\overline{(\tr[\rho^n])^2}-\overline{(\tr[\rho^n Y])^2}\right]/4$ and $\text{Var}[g] = 4\text{Var}[p_\text{denom}] = 1-\overline{(\tr[\rho^n])^2}$. Here, since $f = p_\text{denom}+p_\text{num}-1$ and $g = 2p_\text{denom}-1$ provides $\Delta f = \Delta p_\text{denom}+\Delta p_\text{num}$ and $\Delta g = 2\Delta p_\text{denom}$, we can calculate $\text{E}[\Delta f \Delta g]$ as follows:
\begin{equation}
\text{E}[\Delta f \Delta g] = \text{E}\left[2(\Delta p_\text{denom})^2 + 2\Delta p_\text{num}\Delta p_\text{denom}\right] = 2\text{E}\left[(\Delta p_\text{denom})^2\right] = 2\text{Var}[p_\text{denom}]= \frac{1-\overline{(\text{Tr}[\rho^n])^2}}{2},
\end{equation}
where we have used $\text{E}[\Delta p_\text{num}\Delta p_\text{denom}]=0$, which comes from the above assumption of no correlation between $p_\text{num}$ and $p_\text{denom}$.


\section{Relationship between systematic errors and R\'enyi entropy}\label{sec:Renyi}
In this section, we discuss the relationship between the systematic error $x - x_\text{e}$ and the R\'enyi entropy of the error states under a protocol similar to the one in the main text but for general incoherent noise \cite{koczor2020exponential}. After the time evolution in the magnetic field  with noise for the interaction time $t$ at the $i$th experimental run, the state is described as
\begin{equation}
\rho_i(t)= \mathcal{E}_i \left[e^{-iHt}\rho(0)e^{iHt}\right] = [1-\epsilon_i(t)]e^{-iHt}\rho(0)e^{iHt} + \epsilon_i(t) \sum_{k=1}^{2^L-1}p_k(t)\ket{\psi_k(t)}\bra{\psi_k(t)}
\end{equation}
where $\epsilon_i(t)$ is the error rate at the $i$th experimental run and $\ket{\psi_k(t)}$ is an error state that forms an orthonormal basis set with $e^{-iHt}\ket{\text{GHZ}}$ with $p_k(t)$ being the corresponding probability satisfying $\sum_{k=1}^{2^L-1}p_k(t) = 1$. For error-mitigated quantum metrology, we calculate $\text{Tr}\left[\overline{\rho(t)^n}\right]$ and $\text{Tr}\left[\overline{\rho(t)^n}P_y\right]$ as follows:
\begin{align}
\text{Tr}\left[\overline{\rho(t)^n}\right] &= \overline{[1-\epsilon(t)]^n} + \overline{\epsilon(t)^n}\sum_{k=1}^{2^L-1}[p_k(t)]^n \equiv \overline{[1-\epsilon(t)]^n}+ \overline{\epsilon(t)^n}
\|\mathbf{p}(t)\|_n^n, \\
\text{Tr}\left[\overline{\rho(t)^n} P_y\right]
&
\simeq  \frac{1}{2}\left\{\overline{[1-\epsilon(t)]^n} + \overline{\epsilon(t)^n}[p_\perp(t)]^n\right\} + \frac{1}{2}\left\{\overline{[1-\epsilon(t)]^n}- \overline{\epsilon(t)^n}[p_\perp(t)]^n\right\}L\omega t,
\end{align}
where $\|\mathbf{p}(t)\|_n^n = \sum_{k=1}^{2^L-1}[p_k(t)]^n$ with $\|\mathbf{p}(t)\|_n$ being the n-norm of the probability vector $\mathbf{p}(t) = \left[p_1(t), ..., p_{2^L-1}(t)\right]$, and $p_\perp(t)$ is one of the probabilities in $\{p_k(t)\}_{k=1}^{2^L-1}$ that corresponds to the state $e^{-iHt}\ket{\text{GHZ}_\perp}$ with $\ket{\text{GHZ}_\perp}=(\ket{00...0}-\ket{11...1})/\sqrt{2}$. From these quantities, we obtain the error-mitigated probability,
\begin{align}
\frac{\text{Tr}\left[\overline{\rho(t)^n}
P_y\right]}{\text{Tr}\left[\overline{\rho(t)^n}\right]} &=x + y\omega, \\
x &= \frac{1}{2}\frac{1+E(t)[p_\perp(t)]^n}{1+E(t)\|\mathbf{p}(t)\|_n^n}, \\
y &= \frac{1}{2}\frac{1-E(t)[p_\perp(t)]^n}{1+E(t)\|\mathbf{p}(t)\|_n^n}Lt, 
\end{align}
where $E(t) =\overline{\epsilon(t)^n}/\overline{[1-\epsilon(t)]^n}$. Let $x_\text{e}$ be the $x$ calculated from the estimated error rate $\epsilon_\text{e}$. Then, the difference between the actual and estimated values, $|x-x_{\text{e}}|$, is 
\begin{align}
|x-x_\text{e}| &=\frac{1}{2}\left|\frac{1+E(t)[p_\perp(t)]^n}{1+E(t)\|\mathbf{p}(t)\|_n^n} - \frac{1}{2}\frac{1+E_\text{e}(t)[p_\perp(t)]^n}{1+E_\text{e}(t)\|\mathbf{p}(t)\|_n^n}\right|\\
&=\|\mathbf{p}(t)\|_n^n\frac{\left|E_\text{e}(t)-E(t)\right|\left\{1-\left[\frac{p_\perp(t)}{\|\mathbf{p}(t)\|_n}\right]^n\right\}}{2[1+E(t)\|\mathbf{p}(t)\|_n^n][1+E_\text{e}(t)\|\mathbf{p}(t)\|_n^n]}, \label{eq:general_bias}
\end{align}
where $E_\text{e}(t) =\overline{\epsilon_\text{e}(t)^n}/\overline{[1-\epsilon_\text{e}(t)]^n}$.
This expression clarifies the condition for $x-x_\text{e}=0$ \textit{i.e.}~no systematic errors: $E=E_\text{e}$, which means the correct estimation of error, or $p_\perp(t)=\|\mathbf{p}(t)\|_n$, which means that the error state is only $e^{-iHt}\ket{\text{GHZ}_\perp}$. For example, the latter condition, $p_\perp(t)=\|\mathbf{p}(t)\|_n$, is satisfied with separable initial states of a two-state system and global or local dephasing noise with initial GHZ states. We also find that the $L$ dependence of $|x-x_\text{e}|$ is  determined by that of $\|\mathbf{p}(t)\|_n^n\left|E_\text{e}(t)-E(t)\right|$, so it is roughly determined by that of $\overline{\epsilon(t)^n}\|\mathbf{p}(t)\|_n^n$. Here, $\overline{\epsilon(t)^n}\|\mathbf{p}(t)\|_n^n$ is related to the R\'enyi entropy of the error states \cite{koczor2020exponential},
\begin{equation}
\overline{\epsilon(t)^n}\|\mathbf{p}(t)\|_n^n = e^{-(n-1)H_n},
\end{equation}
where
\begin{equation}
H_n= \frac{1}{1-n}\ln\left[\overline{\epsilon(t)^n}\sum_{k=1}^{2^L-1}[p_k(t)]^n\right] = \frac{n}{1-n}\ln\left\{\left(\overline{\epsilon(t)^n}\right)^{1/n}\|\mathbf{p}(t)\|_n\right\} \label{eq:Renyi}
\end{equation}
is the R\'enyi entropy of the error states. Therefore, the performance of our protocol in the presence of incoherent noise depends on the $L$ dependence of the R\'enyi entropy of the error states. We should remark here that the local amplitude damping in the main text is coherent noise and the above discussion cannot be applied. Nevertheless, the $L$ dependence of $|x-x_\text{e}|$ can be roughly interpreted as the R\'enyi entropy of the error states, as indicated in the next section.
\section{Details of the calculations and discussions for the main text}\label{sec:LocalAD}
Here, we provide details of the calculations and discussions for the main text.
\subsection{Derivation of $\braket{P_y}_\text{mit}$ and $\delta^2\omega$}
The density matrix at the $i$th experimental run after interacting with the target magnetic field with noise is described as

\begin{equation}
\rho_i(t)= \mathcal{A}_i^{(L)} \circ...\circ \mathcal{A}^{(1)}_i [e^{-iHt}\rho(0)e^{iHt}],
\label{Eq:ithstate}
\end{equation}
where $\mathcal{A}_i$ denotes the noise channel of the local amplitude damping acting on the $j$th qubit defined by 
\begin{align}
\mathcal{A}_i^{(j)} [\rho^{(j)}]&= K_{i,1} \rho^{(j)} \bigl[K_{i,1 }\bigr]^\dag+  K_{i,2} \rho^{(j)} \bigl[K_{i,2}\bigr]^\dag, \\
K_{i,1}&= \begin{pmatrix}  
1 & 0\\
0 & \sqrt{1-\epsilon_i(t)}
\end{pmatrix},
K_{i,2}= \begin{pmatrix}  
0 & \sqrt{\epsilon_i(t)}\\
0 & 0
\end{pmatrix},
\label{Eq:amplitude}
\end{align}
with $\epsilon_i(t)$ being the error rate at the $i$th experimental run and $\rho^{(j)}$ being a single-qubit state of the $j$th qubit.
Equation \eqref{Eq:ithstate} provides the following expression for the density matrix:
\begin{align}
\rho_i(t)&= \frac{1}{2}\left[1-\epsilon_i(t)\right]^L\ket{1...1}\bra{1...1}+\frac{1}{2}\left\{1+\left[\epsilon_i(t)\right]^L\right\}\ket{0...0}\bra{0...0} \notag \\
&+\frac{1}{2}\left[1-\epsilon_i(t)\right]^{L/2}\left(e^{-iL\omega t}\ket{1...1}\bra{0...0}+e^{iL\omega t}\ket{0...0}\bra{1...1}\right) 
+ \frac{1}{2}\sum_{k=1}^{2^L-2}p_{k,i}(t)\ket{\psi_k(t)}\bra{\psi_k(t)} \\
&= \frac{1}{2}\left[\lambda_{+,i}(t)\ket{\lambda_{+,i}(t)}\bra{\lambda_{+,i}(t)}+\lambda_{-,i}(t)\ket{\lambda_{-,i}(t)}\bra{\lambda_{-,i}(t)}\right] +\frac{1}{2}\sum_{k=1}^{2^L-2}p_{k,i}(t)\ket{\psi_{k}(t)}\bra{\psi_k(t)}, \label{eq:rho_iAmp}
\end{align}
where  
\begin{equation}
\lambda_{\pm,i}(t) = \frac{1+[\epsilon_i(t)]^L+[1-\epsilon_i(t)]^L\pm \sqrt{\{1+[\epsilon_i(t)]^L-[1-\epsilon_i(t)]^L\}^2+4[1-\epsilon_i(t)]^L}}{2}
\end{equation}
are eigenvalues and
\begin{align}
\ket{\lambda_{+,i}(t)} &= \cos\theta_i(t) e^{-iL\omega t/2}\ket{1...1}+\sin\theta_i(t) e^{iL\omega t/2}\ket{0...0}, \\
\ket{\lambda_{-,i}(t)} &= -\sin\theta_i(t)e^{-iL\omega t/2}\ket{1...1}+\cos\theta_i(t) e^{iL\omega t/2}\ket{0...0}
\end{align}
are the corresponding eigenstates with 
\begin{align}
\cos 2\theta_i(t) &= \frac{[1-\epsilon_i(t)]^L-1-[\epsilon_i(t)]^L}{\sqrt{\{1+[\epsilon_i(t)]^L-[1-\epsilon_i(t)]^L\}^2+4[1-\epsilon_i(t)]^L}}, \\
\sin 2\theta_i(t) &= \frac{2[1-\epsilon_i(t)]^{L/2}}{\sqrt{\{1+[\epsilon_i(t)]^L-[1-\epsilon_i(t)]^L\}^2+4[1-\epsilon_i(t)]^L}}.
\end{align}
Here, $\ket{\psi_k(t)} = e^{-iHt}\ket{\psi_k}$ with $\ket{\psi_k}$ being a computational basis that is orthogonal to $\ket{0...0}$ and $\ket{1...1}$ (e.g.~$\ket{010...0}, \ket{011...0}$), and $p_{k,i}(t) = [\epsilon_i(t)]^k[1-\epsilon_i(t)]^{L-k}$ is the corresponding probability with $k$ being the number of 0s in $\ket{\psi_k}$.

To calculate the error-mitigated expectation value, we calculate  $\tr\left[\overline{\rho(t)^n}\right]$, $\tr\left[\overline{\rho(t)^n}P_y\right]$, and $\tr\left[\overline{\rho(t)^n}Y\right]$ as follows:
\begin{align}
\tr\left[\overline{\rho(t)^n}\right] &=  \frac{1}{2^n}\left[\overline{\lambda_{+}(t)^n}+\overline{\lambda_{-}(t)^n} +\sum_{k=1}^{L-1}\binom{L}{k}\overline{\epsilon(t)^{nk}[1-\epsilon(t)]^{n(L-k)}}\right], \label{eq:Trrho^n} \\
\tr\left[\overline{\rho(t)^n} P_y\right] &\simeq \frac{1}{2}\frac{\overline{\lambda_{+}(t)^n}+\overline{\lambda_{-}(t)^n}}{2^n} + \frac{1}{2}\frac{\overline{\lambda_{+}(t)^n\sin 2\theta(t)}-\overline{\lambda_{-}(t)^n\sin 2\theta(t)}}{2^n} L\omega t, \label{eq:Trrho^nP_y} \\
\text{Tr}\left[\overline{\rho(t)^n} Y\right] &= 2\tr\left[\overline{\rho(t)^n} P_y\right] -\tr\left[\overline{\rho(t)^n}\right].
\end{align}
These quantities provide an error-mitigated probability as 
\begin{align}
\braket{P_y}_\text{mit} &= \frac{\tr[\overline{\rho(t)^n} P_y]}{\tr[\overline{\rho(t)^n}]} = x+y\omega,
\end{align}
where
\begin{align}
x &= \frac{1}{2}\frac{\overline{\lambda_{+}(t)^n}+\overline{\lambda_{-}(t)^n}}{\overline{\lambda_{+}(t)^n}+\overline{\lambda_{-}(t)^n} +\sum_{k=1}^{L-1}\binom{L}{k}\overline{\epsilon(t)^{nk}[1-\epsilon(t)]^{n(L-k)}}}, \label{eq:x_amp} \\
y &= \frac{1}{2}\frac{\overline{\lambda_{+}(t)^n\sin 2\theta(t)}-\overline{\lambda_{-}(t)^n\sin 2\theta(t)}}{\overline{\lambda_{+}(t)^n}+\overline{\lambda_{-}(t)^n} +\sum_{k=1}^{L-1}\binom{L}{k}\overline{\epsilon(t)^{nk}[1-\epsilon(t)]^{n(L-k)}}}Lt \label{eq:y_amp}.
\end{align}
The estimation uncertainty in our protocol is calculated to be $\delta^2\omega = \left(\text{Var}[\braket{P_y}_\text{mit}]+(x-x_\text{e})^2\right)/y_\text{e}^2$, where $\text{Var}[\braket{P_y}_\text{mit}]$ is the variance of $\braket{P_y}_\text{mit}$ calculated with Eq.~\eqref{eq:VarP_y}. Here, $x_\text{e}$ and $y_\text{e}$ are the estimated values of $x$ and $y$, respectively, and they are calculated using Eqs.~\eqref{eq:x_amp}, \eqref{eq:y_amp} with the error rate at the $i$th experimental run, $\epsilon_{\text{e}}(t) = 1-\exp(-t/T_{1,\text{e}})$ for Markovian noise and $\epsilon_{\text{e}}(t) = 1-\exp(-(t/T_{1,\text{e}})^2)$ for time-inhomogeneous noise with $T_{1,\text{e}}$ being the estimated coherence time. Although we adopt the constant estimated coherence time in the main text, we can easily generalize the formalism for the estimated coherence time that fluctuates from one experimental run to another.

\subsection{$L$-dependence of the systematic error}
Here, we discuss the $L$ dependence of the systematic error by considering $x-x_\text{e}$. Since the error rate at the 'optimal' time $\epsilon(t_\text{e}^\text{opt})$ is the order of $L^{-1}$, we will only consider the term of unity, $[1-\epsilon(t_\text{e}^\text{opt})]^L$, and $\epsilon(t_\text{e}^\text{opt})[1-\epsilon(t_\text{e}^\text{opt})]^{L-1}$ and neglect higher order terms of $\epsilon(t_\text{e}^\text{opt})$. Accordingly, we obtain $\lambda_{+,i}(t_\text{e}^\text{opt}) \simeq 1+[1-\epsilon(t_\text{e}^\text{opt})]^L$, $\lambda_{-,i}(t_\text{e}^\text{opt})\simeq 0$ and $\sum_{k=1}^{L-1}\binom{L}{k}[\epsilon_i(t_\text{e}^\text{opt})^{nk}[1-\epsilon_i(t_\text{e}^\text{opt})]^{n(L-k)}\simeq L\epsilon_i(t_\text{e}^\text{opt})^n[1-\epsilon_i(t_\text{e}^\text{opt})]^{n(L-1)}$. Inserting these quantities into Eq.~\eqref{eq:x_amp} yields
\begin{equation}
x \simeq \frac{1}{2}\frac{\overline{\{1+[1-\epsilon(t_\text{e}^\text{opt})]^L\}]^n}}{\overline{\{1+[1-\epsilon(t_\text{e}^\text{opt})]^L\}^n} +L\overline{\epsilon(t_\text{e}^\text{opt})^{n}[1-\epsilon(t_\text{e}^\text{opt})]^{n(L-1)}}}.
\end{equation}
With the approximation above, the systematic error becomes
\begin{align}
& |x-x_\text{e}| \notag \\
&\simeq \frac{1}{2}L\frac{\left|\left\{\overline{[1+[1-\epsilon(t_\text{e}^\text{opt})]^L]^n}\right\}\left\{\overline{\epsilon_\text{e}(t_\text{e}^\text{opt})^n[1-\epsilon_\text{e}(t_\text{e}^\text{opt})]^{n(L-1)}}\right\}-\left\{\overline{[1+[1-\epsilon_\text{e}(t_\text{e}^\text{opt})]^L]^n}\right\}\left\{\overline{\epsilon(t_\text{e}^\text{opt})^n][1-\epsilon(t_\text{e}^\text{opt})]^{n(L-1)}}\right\}\right|}{\left\{\overline{[1+[1-\epsilon(t_\text{e}^\text{opt})]^L]^n} + L\overline{\epsilon(t_\text{e}^\text{opt})^n[1-\epsilon(t_\text{e}^\text{opt})]^{n(L-1)}}\right\}\left\{\overline{[1+[1-\epsilon_\text{e}(t_\text{e}^\text{opt})]^L]^n} + L\overline{\epsilon_\text{e}(t_\text{e}^\text{opt})^n[1-\epsilon_\text{e}(t_\text{e}^\text{opt})]^{n(L-1)}}\right\}} \notag\\
&\sim L^{-(n-1)}. \label{eq:amp_bias}
\end{align}
To derive this scaling, we used the following idea. Since we numerically found $t^\text{opt}_\text{e} \sim L^{-1}$ and $t^\text{opt}_\text{e} \sim L^{-1/2}$ for Markovian and time-inhomogeneous noise, respectively, $[1-\epsilon_i(t_\text{e}^\text{opt})]^L = \exp[-Lt_\text{e}^\text{opt}/T_{1,i}] \sim \text{const.}$ for Markovian noise and $[1-\epsilon_i(t_\text{e}^\text{opt})]^L = \exp[-L(t_\text{e}^\text{opt}/T_{1,i})^2] \sim \text{const.}$ for time-inhomogeneous noise. Here, for large $L$, $\epsilon_i(t_\text{e}^\text{opt}) = 1-\exp(-t_\text{e}^\text{opt}/T_{1,i}) \simeq t_\text{e}^\text{opt}/T_{1,i} \sim L^{-1}$ for Markovian noise and $\epsilon_i(t_\text{e}^\text{opt}) = 1-\exp[-(t_\text{e}^\text{opt}/T_{1,i})^2] \simeq (t_\text{e}^\text{opt}/T_{1,i})^2 \sim L^{-1}$ for time-inhomogeneous noise. Therefore, $|x-x_\text{e}|$ scales as $L^{-(n-1)}$ and decreases as $L$ increases. We can understand the scaling in Eq.~\eqref{eq:amp_bias} in terms of the R\'enyi entropy of the error states, although the discussion in Sec.~\ref{sec:Renyi} cannot be directly applied because the local amplitude damping involves coherent errors. The R\'enyi entropy of the error states $\ket{\psi_k(t)}$ with the approximation above is roughly calculated using Eqs.~\eqref{eq:Renyi} and \eqref{eq:rho_iAmp} as
\begin{equation}
H_n=  \frac{1}{1-n}\ln\left[L\overline{\epsilon(t_\text{e}^\text{opt})^n[1-\epsilon(t_\text{e}^\text{opt})^n]^{L-1}}\right] \sim \frac{1}{1-n}\ln[(L^{-n+1})] = \ln(L),
\end{equation}
Thus, it increases as $L$ increases. This rough approximation reproduces the above exact result of the scaling in Eq.~\eqref{eq:amp_bias}.

\subsection{Detailed numerical procedure}
The detailed numerical procedure to calculate $\delta^2\omega$ for demonstration is as follows.
First, we need to determine the interaction time to minimize the variance as in the standard quantum metrology. 
For this purpose, using the \textit{estimated} coherence time $T_\text{1,e}$, we calculate the variance of $\omega$, $\delta^2\omega_\text{e} = \text{Var}[p_\text{e}]/(y_\text{e})^2$ and numerically find a 'pseudo-optimal' interaction time $t = t_\text{e}^\text{opt}$ that minimizes $\delta^2\omega_\text{e}$.
Second, we implement $N_\text{samp}$ numerical runs with $t=t_\text{e}^{\text{opt}}$ under the actual coherence time to compute $\braket{P_y}_\text{mit}= \tr\left[\overline{\rho(t)^n} P_y\right]/\tr\left[\overline{\rho(t)^n}\right]\equiv p = x+y\omega$.
Separately, we calculate $\braket{P_y}_\text{mit}$ with estimated coherence time $T_\text{1,e}$ and interpret this as $ p_\text{e} = x_\text{e}+y_\text{e}\omega$ [see Eq.~\eqref{eq:P_e}].
Finally, we calculate $\delta^2\omega$ using Eq.~\eqref{eq:delta^2omega}.
Note that the cost of generating $2n$ more states is considered as $N_\text{samp}/(2n)$ for fair comparison among different values of $n$.
\subsection{$n$ dependence of $\delta^2\omega$}
\begin{figure}[t]
    \centering
    \includegraphics[width= 0.6\linewidth]{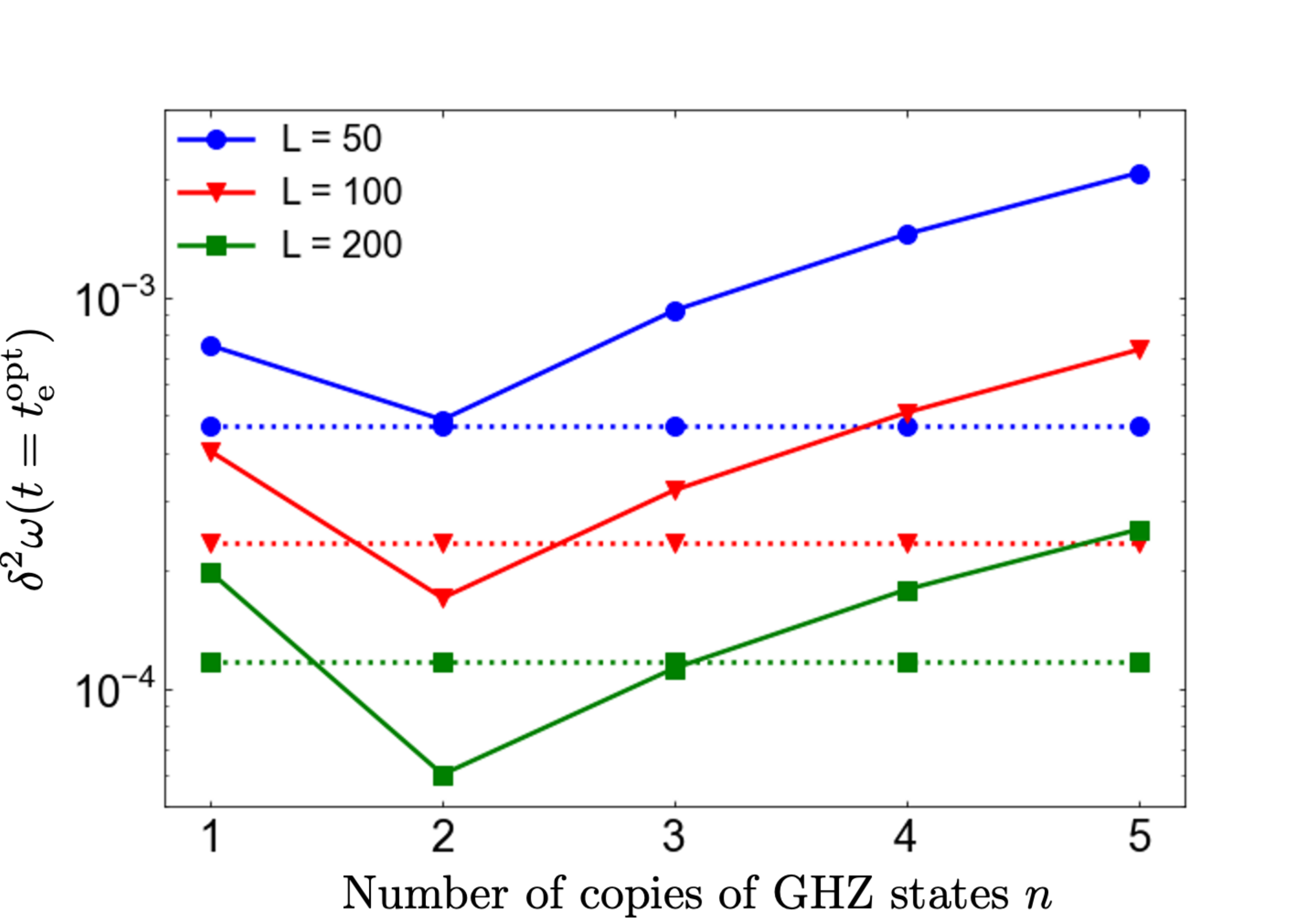}
    \caption{Estimation uncertainty $\delta^2\omega$ for time-inhomogeneous local amplitude damping on GHZ initial states as a function of the number of copies of GHZ states $n$ for some fixed values of the number of qubits $L$. Each dotted line shows $\delta^2\omega$ with $L$ separable states corresponding to that $L$, which indicates the threshold for quantum enhancement. For $n=2$, we can obtain quantum-enhanced $\delta^2\omega$ for $L>50$ in the present parameter setting.}
   \label{fig:GHZNdep}
\end{figure}
In the main text, we discuss the scaling of $\delta^2\omega$ with respect to $L$, which is a typical figure of merit in the theory of conventional quantum metrology.
Here, we investigate what value of $n$ is required to recover the quantum-enhanced $\delta^2\omega$ for fixed $L$ in the presence of fluctuating noise with our error-mitigated quantum metrology.

Figure \ref{fig:GHZNdep} shows $\delta^2\omega$ for $L$ separable initial states as dotted lines and one for $L$-qubit GHZ states as solid lines.
The quantum enhancement is considered to be recovered when $\delta^2\omega$ with entangled states becomes smaller than that with the separable states for the value of $L$.
For $L=50,\ 100,\ 200$, $n=2$ is always optimal in Fig.~\ref{fig:GHZNdep} as discussed in the main text.
For $n=2$, we can obtain quantum-enhanced $\delta^2\omega$ for $L>50$ in the present parameter setting.
Thus, we expect practical usefulness of our protocol even in the current technology.

\subsection{Suitable hardware for the implementation}
We here discuss suitable hardware for the implementation of our protocol for magnetic field sensing by using superconducting circuits.
In our scheme, there are mainly two parts in quantum circuits: the quantum metrology part and the QEM part [see Fig.~\ref{fig:Numerics}(a) in the main text].
The first part is to prepare GHZ states and to expose these states to the magnetic fields; we can use superconducting flux qubits (FQs) since it couples strongly with magnetic fields \cite{bal2012ultrasensitive,budoyo2020electron,toida2019electron}. 
The other part is to implement the purification-based error mitigation; we can use superconducting transmon qubits (TQs) since it has a long coherence time and good controllability.
Indirect coupling between FQs and TQs via superconducting resonators \cite{stern2014flux,rigetti2012superconducting} allows to transfer the states from FQs to TQs.

\section{Local and generalized amplitude damping on GHZ initial states} \label{sec:geneAD}

In the present section, we consider more complicated and general noise model than that in the main text: generalized amplitude damping \cite{wen-jie2017protecting} on GHZ initial states.
The density matrix at the $i$th experimental run after interacting with the target magnetic field with local generalized amplitude damping is described as
\begin{equation}
\rho_i(t)= \mathcal{A}_i^{(L)} \circ...\circ \mathcal{A}^{(1)}_i [e^{-iHt}\rho(0)e^{iHt}],
\label{Eq:geneithstate}
\end{equation}
where $\mathcal{A}_i$ denotes the noise channel on the $j$th qubit state $\rho^{(j)}$ defined by \cite{wen-jie2017protecting}
\begin{align}
\mathcal{A}_i^{(j)} [\rho^{(j)}]&= K_{i,1} \rho^{(j)} \bigl[K_{i,1 }\bigr]^\dag+  K_{i,2} \rho^{(j)} \bigl[K_{i,2}\bigr]^\dag + K_{i,3} \rho^{(j)} \bigl[K_{i,3}\bigr]^\dag +  K_{i,4} \rho^{(j)} \bigl[K_{i,4}\bigr]^\dag
\end{align}
where 
\begin{align}
K_{i,1}&= \sqrt{p_i(t)}\begin{pmatrix}  
1 & 0\\
0 & \sqrt{1-\epsilon_i(t)}
\end{pmatrix},
K_{i,2}= \sqrt{p_i(t)}\begin{pmatrix}  
0 & \sqrt{\epsilon_i(t)}\\
0 & 0
\end{pmatrix}, \notag \\
K_{i,3}&= \sqrt{1-p_i(t)}\begin{pmatrix}  
\sqrt{1-\epsilon_i(t)} & 0\\
0 & 1
\end{pmatrix},
K_{i,4}= \sqrt{1-p_i(t)}\begin{pmatrix}  
0 & 0\\
\sqrt{\epsilon_i(t)} & 0
\end{pmatrix},
\label{Eq:geneamplitude}
\end{align}
Here, $p_i(t)$ is defined as
\begin{equation}
p_i(t) = \frac{1}{1+e^{-\beta_i}},
\end{equation}
where $\beta_i(t)$ is the dimensionless inverse temperature including bare frequency at the $i$th experimental run. 
Here, $\epsilon_i(t)$ is the error rate at the $i$th experimental run. $\epsilon_{i}(t) = 1-\exp(-t/T_{1,i})$ for Markovian noise and $\epsilon_{i}(t) = 1-\exp(-(t/T_{1,i})^2)$ for time-inhomogeneous noise, where 
\begin{equation}
T_{1,i} = \frac{1}{(2N_i+1)\gamma_i}
\end{equation}
with $N = (e^{\beta_i}-1)^{-1}$ and $\gamma_i$ being the spontaneous emission rate at the $i$th experimental run.
\begin{figure}
    \centering
    \includegraphics[width= \linewidth]{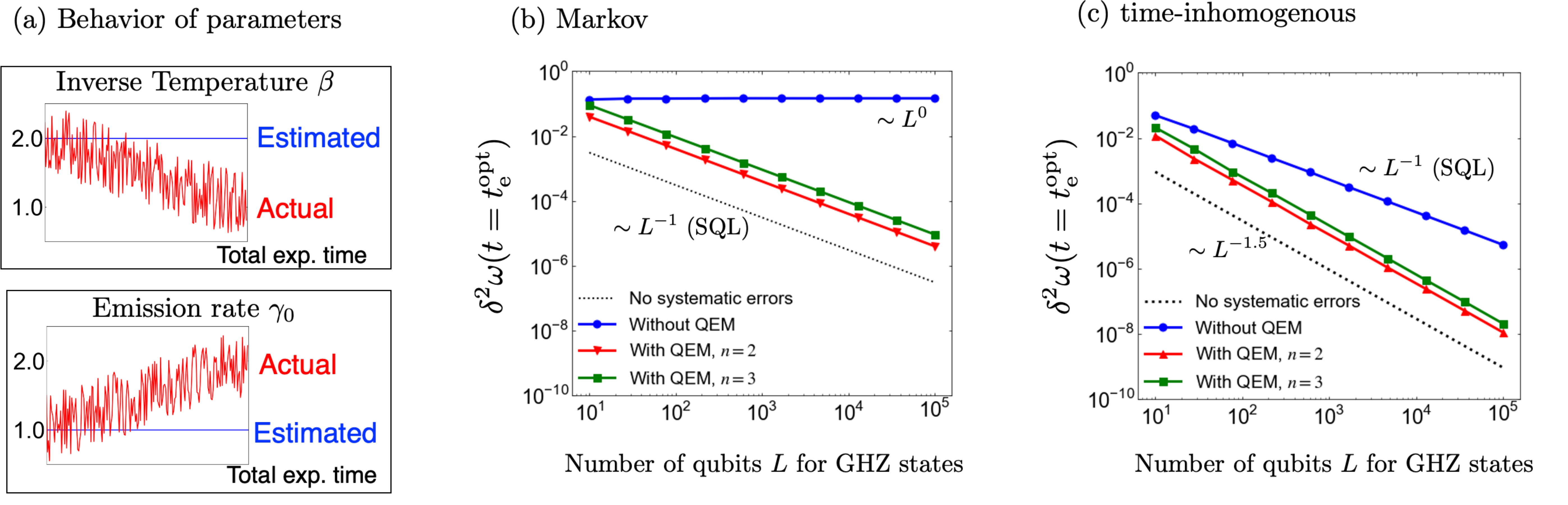}
    \caption{(a) Behaviors of estimated and actual inverse temperature $\beta$ and emission rate $\gamma_0$ in the numerical simulation. We show the estimation uncertainty $\delta^2\omega$ for (b) Markovian and (c) time-inhomogeneous local amplitude damping at finite temperatures on GHZ initial states. Although the ideal scaling is realized by the unbiased estimator with the correct estimation of the noise model (black dotted line), a biased estimator with an incorrect estimation of the noise model leads to systematic errors and spoils the scaling (blue line with circles). Our protocol mitigates systematic error and recovers the ideal scaling (red and green lines with triangles and squares, respectively).}
   \label{fig:geneAmp}
\end{figure}
Equation \eqref{Eq:geneithstate} provides the following expression for the density matrix:
\begin{align}
\rho_i&= \frac{1}{2}\left\{a_i(t)\ket{1...1}\bra{1...1}+b_i(t)[e^{-iL\omega t}\ket{1...1}\bra{0...0}+e^{iL\omega t}\ket{0...0}\bra{1...1}]+c_i(t)\ket{0...0}\bra{0...0}\right\} \\ \notag 
&+ \frac{1}{2}\sum_{k=1}^{2^L-2}A_{k,i}(t)\ket{\psi_k(t)}\bra{\psi_k(t)} \\
&= \frac{1}{2}\left[\lambda_{+,i}(t)\ket{\lambda_{+,i}(t)}\bra{\lambda_{+,i}(t)}+\lambda_{-,i}(t)\ket{\lambda_{-,i}(t)}\bra{\lambda_{-,i}(t)}\right] +\frac{1}{2}\sum_{k=1}^{2^L-2}A_{k,i}(t)\ket{\psi_{k}(t)}\bra{\psi_k(t)}, \label{eq:generho_iAmp}
\end{align}
where  
\begin{align}
a_i(t) &= [1-p_i(t)\epsilon_i(t)]^L+[1-p_i(t)]^L\epsilon_i(t)^L, \\ 
b_i(t) &= [1-\epsilon_i(t)]^{L/2}, \\
c_i(t) &= [1-(1-p_i(t))\epsilon_i]^L+p_i(t)^L\epsilon_i(t)^L, \\
A_{k,i}(t) &= p_i(t)^k\epsilon_i(t)^k[1-p_i(t)\epsilon_i(t)]^{L-k}+\epsilon^{L-k}[1-p_i(t)]^{L-k}[1-(1-p_i(t))\epsilon_i(t)]^{k},
\end{align}
$\ket{\psi_k(t)} = e^{-iHt}\ket{\psi_k}$ with $\ket{\psi_k}$ being a computational basis that is orthogonal to $\ket{0...0}$ and $\ket{1...1}$ (e.g.~$\ket{010...0}, \ket{011...0}$) with $k$ being the number of 0s in $\ket{\psi_k}$, and
\begin{equation}
\lambda_{\pm,i}(t) =\frac{1}{2}\left\{a_i(t) + c_i(t) \pm \sqrt{[a_i(t)-c_i(t)]^2+4b_i(t)^2}\right\}
\end{equation}
are eigenvalues. 
The corresponding eigenstates are as follows;
\begin{align}
\ket{\lambda_{+,i}(t)} &= \cos\theta_i(t) e^{-iL\omega t/2}\ket{1...1}+\sin\theta_i(t) e^{iL\omega t/2}\ket{0...0}, \\
\ket{\lambda_{-,i}(t)} &= -\sin\theta_i(t)e^{-iL\omega t/2}\ket{1...1}+\cos\theta_i(t) e^{iL\omega t/2}\ket{0...0},
\end{align}
where
\begin{align}
\cos 2\theta_i(t) &= \frac{a_i(t)-c_i(t)}{\sqrt{[a_i(t)-c_i(t)]^2+4b_i(t)^2}}, \  
\sin 2\theta_i(t)=\frac{2b_i(t)}{\sqrt{[a_i(t)-c_i(t)]^2+4b_i(t)^2}}.
\end{align}
With the calculations above, we obtain  
\begin{align}
\tr\left[\overline{\rho(t)^n}\right] &=  \frac{1}{2^n}\left[\overline{\lambda_{+}(t)^n}+\overline{\lambda_{-}(t)^n} +\sum_{k=1}^{L-1}\binom{L}{k}\overline{A_{k}(t)^n}\right].
\end{align}
The remains of the calculations are the same as that with Sec.~\ref{sec:LocalAD}

Here, we consider the case, where we wrongly estimated constant $\gamma_\text{e} = 1.0$ and $\beta_\text{e} = 2.0$ and the correct ones are linearly drifted as $\gamma_i = 1.0 \text{ to} \ 2.0$ and $\beta_i = 2.0 \text{ to} \ 1.0$ with adding uniformly fluctuating value between $[-0.5. 0.5]$; see Fig.~\ref{fig:geneAmp}(a). The remaining numerical settings are the same as in the main text.
The numerical results for Markovian and time-inhomogeneous cases are shown in Fig.~\ref{fig:geneAmp}(b) and (c) and clearly show that our protocol is also effective for local generalized amplitude damping, which is more complicated than in the main text.


\section{Global depolarizing noise on GHZ initial states for a large R\'enyi entropy}\label{sec:globalDepo}

In Sec.~\ref{sec:Renyi}, we showed that error-mitigated quantum metrology works well for a large R\'enyi entropy. Here, we demonstrate an improvement of $\delta^2\omega$ for the case involving almost the largest R\'enyi entropy: the global depolarizing noise on GHZ initial states. This noise model is rather artificial but it is expected to provide the highest performance in our scheme. The protocol similar to the one in the main text but uses global depolarizing noise on the GHZ initial states as follows: after the time evolution in the magnetic field with noise for the interaction time $t$, the output state at the $i$th experimental run is described as
\begin{align}
\rho_i(t)&= \mathcal{E}\left[e^{-iHt}\rho(0)e^{iHt}\right] = [1-\epsilon_i(t)]e^{-iHt}\rho(0)e^{iHt}+\frac{\epsilon_i(t)}{2^L} I \notag \\
&= \left[1-\epsilon'_i(t)\right]e^{-iHt}\rho(0)e^{iHt}+\epsilon_i'(t)\sum_{k=1}^{2^L-1}p\ket{\psi_k(t)}\bra{\psi_k(t)}, \label{eq:rho_GHZdepo}
\end{align}
where $\epsilon_i(t)$ is the error rate at the $i$th experimental run, $\epsilon'_i(t) = \epsilon_i(t)\left(1-2^{-L}\right)$, $p = \left(2^{L}-1\right)^{-1}$, and $\ket{\psi_k(t)} = e^{-iHt}\ket{\psi_k}$ with $\ket{\psi_k}$ being the state orthogonal to $\ket{\text{GHZ}}$. Note that we have used the fact that the channel of depolarizing noise commutes with the unitary evolution induced by the Hamiltonian $H$. The effect of fluctuating noise is included in $\epsilon_i(t) = 1-\exp(-t/T_i)$ for Markovian noise and $\epsilon_i(t) = 1-\exp(-(t/T_i)^2)$ for time-inhomogeneous noise, where $T_i$ is the coherence time at the $i$th experimental run and fluctuates from one experimental run to another. From the input $2n$ copies of $\rho_i(t)$, we can obtain the error-mitigated expectation value in Eq.~(\ref{Eq:therem1}) by computing the average of the denominators $\tr\left[\overline{\rho(t)^n}\right]$ and numerators $\tr\left[\overline{\rho(t)^n} Y\right]$,
\begin{align}
\tr\left[\overline{\rho(t)^n}\right] &=  \overline{[1-\epsilon'(t)]^n}+\overline{\epsilon'(t)^n}p^{n-1}, \\
\tr\left[\overline{\rho(t)^n} P_y\right]  &\simeq \frac{1}{2}\left\{\overline{[1-\epsilon'(t)]^n}+\overline{\epsilon'(t)^n}p^n\right\} + \frac{1}{2}\left\{\overline{[1-\epsilon'(t)]^n}-\overline{\epsilon'(t)^n}p^n\right\}L\omega t, \\
\text{Tr}\left[\overline{\rho(t)^n} Y\right] &= 2\tr\left[\overline{\rho(t)^n} P_y\right] -\tr\left[\overline{\rho(t)^n}\right].
\end{align}
These expectation values lead to the error-mitigated probability,
\begin{align}
\braket{P_y}_\text{mit} &= \frac{\tr[\overline{\rho(t)^n} P_y]}{\tr[\overline{\rho(t)^n}]} = x+y\omega,
\end{align}
where
\begin{align}
x &= \frac{1}{2}\frac{\overline{[1-\epsilon'(t)]^n}+\overline{\epsilon'(t)^n}p^n}{\overline{[1-\epsilon'(t)]^n}+\overline{\epsilon'(t)^n}p^{n-1}}, \\
y &= \frac{1}{2}\frac{\overline{[1-\epsilon'(t)]^n}-\overline{\epsilon'(t)^n}p^n}{\overline{[1-\epsilon'(t)]^n}+\overline{\epsilon'(t)^n}p^{n-1}}Lt.
\end{align}

\begin{figure}
    \centering
    \includegraphics[width= 0.8\linewidth]{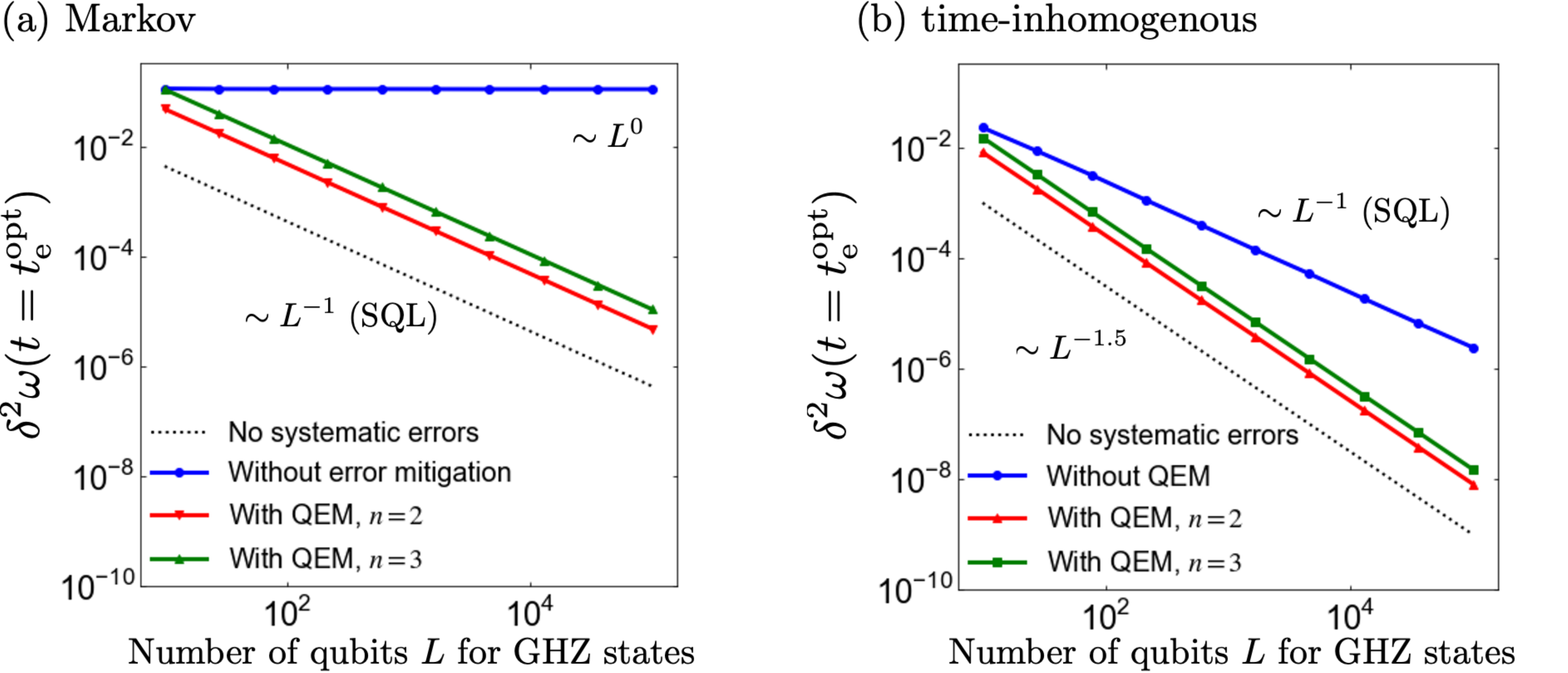}
    \caption{Estimation uncertainty $\delta^2\omega$ for (a) Markovian and (b) time-inhomogeneous global depolarizing noise on GHZ initial states. Although the ideal scaling is realized by the unbiased estimator with correct estimation of the noise model (black dotted line), a biased estimator with an incorrect estimation of the noise model  leads to systematic errors and spoils the scaling (blue line with circles). Our protocol mitigates systematic error and recovers the ideal scaling (red and green lines with triangles and squares, respectively).}
   \label{fig:GHZdepo}
\end{figure}

$\delta^2\omega$ in this protocol is calculated in the same way as in the main text and is shown in Fig.~\ref{fig:GHZdepo}; we have used the same parameters and settings as in the main text. The behavior is the same as that for local amplitude damping noise in the main text. In the present example, the systematic error is exponentially suppressed not only by increasing $n$ but also by increasing $L$. Indeed, $|x- x_{\text{e}}| \sim 2^{-L(n-1)}$, since $\|\mathbf{p}(t)\|_n^n \sim 2^{-L(n-1)}$ as calculated with Eqs.~\eqref{eq:general_bias} and \eqref{eq:rho_GHZdepo}. As shown in Sec.~\ref{sec:Renyi}, this exponential suppression of the systematic error is due to the $L$ dependence of the R\'enyi entropy of the error states, which is calculated to be
\begin{equation}
H_n = \frac{1}{1-n}\ln\left[\overline{\epsilon'(t_\text{e}^\text{opt})^n}\sum_{k=1}^{2^L-1}p^n\right] \sim \ln(2^L-1).
\end{equation}
Intuitively, the number of the error states under global depolarizing noise increases exponentially with $L$, and the R\'enyi entropy of the error states increases. This would be a merit of using our error-mitigated quantum metrology in cases with initial GHZ states.


\section{Separable states of a three-level system for describing the Nitrogen-vacancy center}\label{sec:NV}

So far, we have considered the GHZ state as an initial state because of its potential for entanglement metrology. Here, we turn our attention to a sensor that has been realized with current technology: the nitrogen-vacancy (NV) center \cite{degan2017quantum,barry2020sensitivity}. The NV center is composed of a nearest-neighbor nitrogen atom and a vacancy in diamond and provides a three-level system. Even at room temperature, we can polarize the NV center by illuminating green light, read out the state of the electron spins by measuring the photon emission from the NV center \cite{gruber1997scanning}, and control the NV center by using microwave pulses \cite{jelezko2004observation}. Moreover, the NV center in diamond has a long coherence time (a few milliseconds) even at room temperature \cite{balasubramanian2009ultralong,herbschleb2019ultra}. These properties make it a promising candidate for a highly-sensitive quantum sensor \cite{degan2017quantum,barry2020sensitivity}.

One of the challenges to improving the sensitivity of the diamond-based quantum sensor is decoherence. NV centers are affected by dephasing and energy relaxation. It is known that we can suppress dephasing on the NV center by using dynamical decoupling \cite{de2010universal}. This means that the coherence time of the NV center is ultimately limited by energy relaxation \cite{hein2005entanglement}.

In this section, to show the performance of our protocol for sensing with the NV center, we consider $L$ separable states of a three-level system under depolarizing noise, which corresponds to an energy relaxation in the high-temperature limit.

We consider a system composed of three states, $\ket{\pm 1}, \ket{0}$, where $\ket{0}$ and $\ket{\pm 1}$ are the eigenstates of the Pauli Z operator $\hat{\sigma}_z$, such as $\hat{\sigma}_z\ket{0} = 0$ and $\hat{\sigma}_z\ket{\pm 1} = \pm \ket{1}$. We choose an initial probe state $\rho(0)=(\ket{+}\bra{+})^{\otimes L}$ with $\ket{+} = (\ket{1}+\ket{-1})/\sqrt{2}$. We prepare $2n$ copies of $\rho(0)$ for error-mitigated quantum metrology. We consider the Zeeman Hamiltonian $H= \sum_{j=1}^{L}\omega \hat{\sigma}_z^{(j)}/2$ with a parameter $\omega$ determined by the target field. After the time evolution in the magnetic field with noise for the interaction time $t$, the state at the $i$th experimental run is described as
\begin{equation}
\rho_i(t)= \mathcal{E}_i\left[e^{-iHt}\rho(0)e^{iHt}\right] = [1-\epsilon_i(t)]e^{-iHt}\rho(0)e^{iHt}+\frac{\epsilon_i(t)}{3}I \label{eq:rhoNV}
\end{equation}
with $\epsilon_i(t)$ being the error rate at the $i$th experimental run. The effect of fluctuating noise is included in $\epsilon_i(t) = 1-\exp(-t/T_i)$ for Markovian noise and $\epsilon_i(t) = 1-\exp(-(t/T_i)^2)$ for time-inhomogeneous noise, where $T_i$ is the coherence time at the $i$th experimental run and fluctuates from one experimental run to another. From the input $2n$ copies of $\rho_i(t)$, we can obtain the error-mitigated expectation value in Eq.~(\ref{Eq:therem1}) by computing the average of the denominators $\tr\bigl[\overline{\rho(t)^n}\bigr]$ and numerators $\tr\bigl[\overline{\rho(t)^n} Y\bigr]$ as
\begin{align}
\tr\left[\overline{\rho(t)^n}\right]&= \overline{\left(1-\frac{2}{3}\epsilon(t)\right)^n} +2\overline{\left(\frac{\epsilon(t)}{3}\right)^n}, \\
\tr\left[\overline{\rho(t)^n} P_y\right]  &\simeq \frac{1}{2}\left[\hspace{0.5mm}\overline{\left(1-\frac{2}{3}\epsilon(t)\right)^n} + \overline{\left(\frac{\epsilon(t)}{3}\right)^n}\hspace{0.5mm}\right] + \frac{1}{2}\left[\hspace{0.5mm}\overline{\left(1-\frac{2}{3}\epsilon(t)\right)^n} - \overline{\left(\frac{\epsilon(t)}{3}\right)^n}\hspace{0.5mm}\right]\omega t, \\
\text{Tr}\left[\overline{\rho(t)^n} Y\right] &= 2\tr\left[\overline{\rho(t)^n} P_y\right] -\tr\left[\overline{\rho(t)^n}\right].
\end{align}

These expectation values lead to an error-mitigated probability,
\begin{align}
\braket{P_y}_\text{mit} &= \frac{\tr[\overline{\rho(t)^n} P_y]}{\tr[\overline{\rho(t)^n}]} = x+y\omega,
\end{align}
where
\begin{align}
x &= \frac{1}{2}\frac{\overline{\left[1-\frac{2}{3}\epsilon(t)\right]^n} + \overline{\left[\frac{\epsilon(t)}{3}\right]^n}}{\overline{\left[1-\frac{2}{3}\epsilon(t)\right]^n} +2\overline{\left[\frac{\epsilon(t)}{3}\right]^n}}, \\
y &= \frac{1}{2}\frac{\overline{\left[1-\frac{2}{3}\epsilon(t)\right]^n} - \overline{\left[\frac{\epsilon(t)}{3}\right]^n}}{\overline{\left[1-\frac{2}{3}\epsilon(t)\right]^n} +2\overline{\left[\frac{\epsilon(t)}{3}\right]^n}}Lt.
\end{align}

\begin{figure}[t]
    \centering
    \includegraphics[width= 0.8\linewidth]{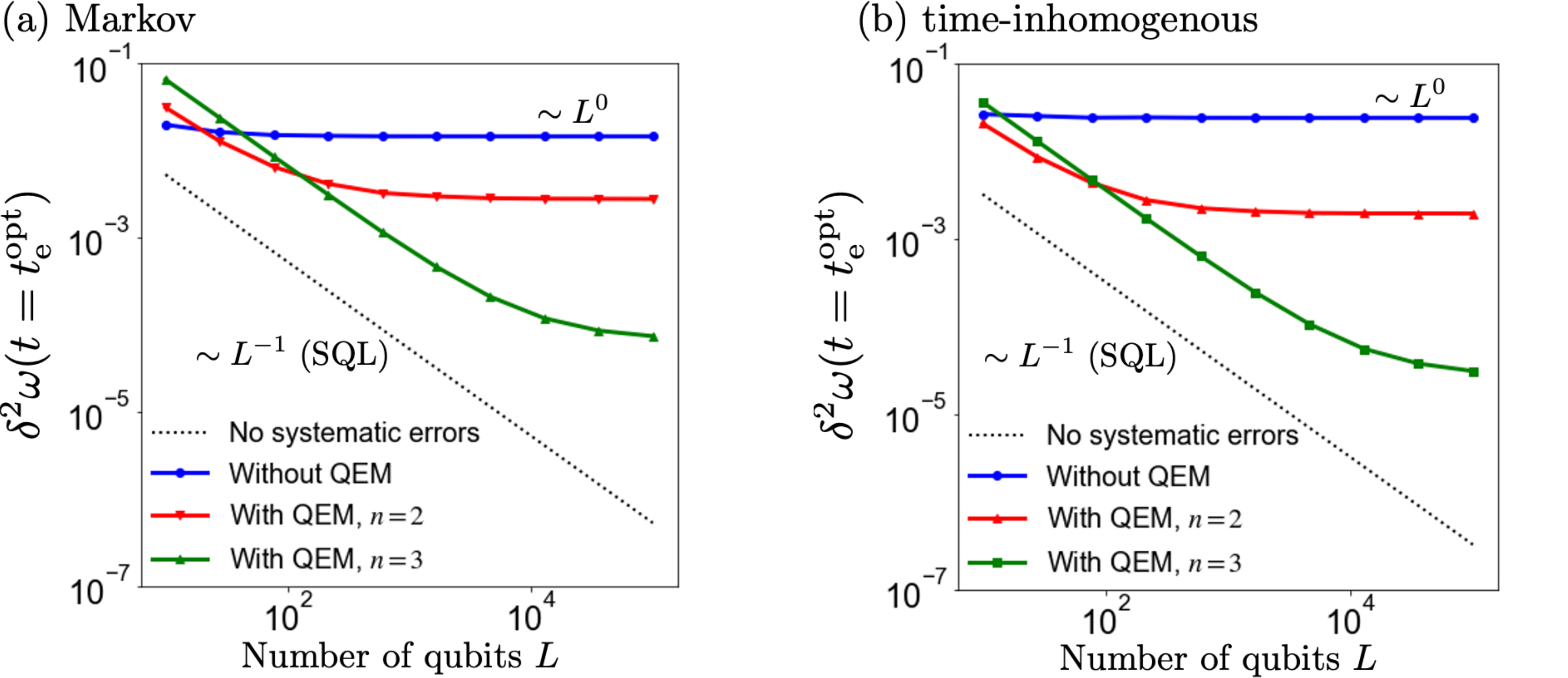}
    \caption{Estimation uncertainty $\delta^2\omega$ for $L$ separable states of a three-level system under (a) Markovian and (b) time-inhomogeneous depolarizing noise. Although the ideal scaling is realized by the unbiased estimator with correct estimation of the noise model (black dotted line), a biased estimator with an incorrect estimation of the noise model  leads to systematic errors and spoils the scaling (blue line with circles). Our protocol mitigates systematic error and recovers the ideal scaling (red and green lines with triangles and squares, respectively).}
   \label{fig:NV}
\end{figure}

The estimation uncertainty in our protocol is calculated in the same way as in the main text and is shown in Fig.~\ref{fig:NV}. We used the same parameters and settings as in the main text; the only difference is that we consider $N_\text{samp}\to LN_\text{samp}$ because we treat $L$-qubit separable states. Although our scheme works well and recovers the scaling of $\delta^2\omega$ in the present case, the behavior of $\delta^2\omega$ is different from the previous case and the case in the main text with initial GHZ states. The qualitative behavior of $\delta^2\omega$ is independent of both Markovian and time-inhomogeneous noise as shown in Fig.~\ref{fig:NV}, since the initial states are separable and not entangled; it is known that the scaling of $\delta^2\omega$ for separable states is limited to SQL \cite{degan2017quantum}. Moreover, when the systematic error is dominant, the optimal $n$ increases as $L$ increases. This is because the systematic error in the present case is independent of $L$: the R\'enyi entropy is also independent of $L$. The systematic error is, therefore, suppressed only by increasing $n$, involving an exponentially increasing sampling cost \cite{koczor2020exponential}. When the statistical error is dominant, i.e., for small $L$, the increased sampling cost degrades $\delta^2\omega$, and therefore, a smaller $n$ is optimal. When the systematic error is dominant, i.e., for large $L$, since increasing $n$ exponentially suppresses the systematic error, a large $n$ is optimal. This is the reason for the crossover behavior between the cases of $n=2$ and $n=3$ observed in Fig.~\ref{fig:NV}.

\end{document}